\documentclass[11pt]{scrartcl}
\usepackage{multirow}
\usepackage{makecell}
\usepackage[ngerman, english]{babel}
\usepackage[latin1]{inputenc}
\usepackage[T1]{fontenc}
\usepackage{graphicx}
\usepackage{url}
\usepackage{marvosym}
\usepackage{lmodern}
\usepackage{comment}
\usepackage[table]{xcolor}
\usepackage{xcolor}
\usepackage{float}
\usepackage{titletoc}

\usepackage[doublespacing]{setspace}

\usepackage{fancyref}
\usepackage{enumitem}	
\usepackage{tabularx}	
\usepackage{longtable}	% Tabellen über mehrere Seiten
\usepackage{booktabs}	% Tabellenlayout
\usepackage{dcolumn}	% In Regressionstabellen Zahlen am Dezimalpunkt zentrieren
\usepackage{threeparttable}
\usepackage{pdflscape} 	% Seiten im Querformat
\usepackage{amsmath, amssymb}
\usepackage{makecell}

\usepackage[official]{eurosym} % Official Euro-Symbol
\usepackage{caption}           % Tabellen- und Figurenüberschriften
\usepackage{subcaption}
\usepackage{ragged2e}
\usepackage{afterpage}

\usepackage[left=2.5cm, right=2.5cm, bottom=2.5cm, top=2.5cm]{geometry} 
\usepackage{graphicx} %Grafiken einbinden
\usepackage{mathtools} %Mathematik
\usepackage{mathpazo} %Mathematik
\usepackage{array}  %Column width
\usepackage[title]{appendix}

\usepackage[all]{nowidow}

\newcolumntype{C}[1]{>{\centering\let\newline\\\arraybackslash\hspace{0pt}}m{#1}}
\newcolumntype{L}[1]{>{\raggedright\let\newline\\\arraybackslash\hspace{0pt}}m{#1}}
\newcolumntype{R}[1]{>{\raggedleft\let\newline\\\arraybackslash\hspace{0pt}}m{#1}}
\newcolumntype{d}[1]{D{.}{.}{#1}}

\usepackage[pdftex,bookmarks=true, colorlinks=false, pdfborder={0 0 0}]{hyperref} 	

\usepackage{apacite}   %Unterschied: apacite und apalike \usepackage{apalike}
\usepackage{dcolumn}

\begin{document}

%%%
% Title page:

\title{A 22 percent increase \\ in the German minimum wage: \\ nothing crazy! \vspace{0.7cm}}

\begin{onehalfspacing}

\author{Mario Bossler%
\thanks{TH Nuremberg Georg Simon Ohm (TH Nuremberg), Institute for Employment Research (IAB), Institute of Labor Economics (IZA), and Labor and Socio-Economic Research Center of the University of Erlangen-Nuremberg (LASER). Address: TH Nuremberg, Kesslerplatz 12, 90489 Nuremberg, Germany, \Letter~\ \texttt{mario.bossler@th-nuernberg.de}.}\vspace{-0.3cm}
\\
\small{\emph {TH Nuremberg, IAB, IZA, LASER}}
\and
Lars Chittka%
\thanks{German Statistical Office (Destatis). Address: 
Gustav-Stresemann-Ring 11, 65189 Wiesbaden, Germany, 
\Letter~\ \texttt{Lars.Chittka@destatis.de}.}\vspace{-0.3cm}
\\
\small{\emph {Destatis}}
\and
Thorsten Schank%
\thanks{Johannes Gutenberg-University Mainz (JGU Mainz), Institute of Labor Economics (IZA), and Labor and Socio-Economic Research Center of the University of Erlangen-Nuremberg (LASER). Address: JGU Mainz, Jakob-Welder-Weg 4, 55128 Mainz, \Letter~\ \texttt{schank@uni-mainz.de}.
\newline
We gratefully acknowledge comments from David Card whose assessment ``nothing crazy'' (in terms of potentially detrimental effects of the minimum wage) led us to choose the title, as well as comments from Michael Oberfichtner, Martin Popp, and those received at the 4th Scientific Workshop of the German Minimum Wage Commission in Berlin and from seminars at the IAB Nuremberg, JMU Wuerzburg, and JGU Mainz.}\vspace{-0.3cm}
\\
\small{\emph {JGU Mainz, IZA, LASER}} \vspace{0.5cm}
}

\date{November 2024}
\maketitle

\vspace{-0.5cm}
\begin{abstract}
\noindent We present the first empirical evidence on the 22 percent increase in the German minimum wage, implemented in 2022, raising it from \euro{9.82} to \euro{10.45} in July and to \euro{12} in October. Leveraging the German Earnings Survey, a large and novel data source comprising around 8 million employee-level observations reported by employers each month, we apply a difference-in-difference-in-differences approach to analyze the policy's impact on hourly wages, monthly earnings, employment, and working hours. Our findings reveal significant positive effects on wages, affirming the policy's intended benefits for low-wage workers. Interestingly, we identify a negative effect on working hours, mainly driven by minijobbers. The hours effect results in an implied labor demand elasticity in terms of the employment volume of $-0.2$ which only partially offsets the monthly wage gains. We neither observe a negative effect on the individual's employment retention nor the regional employment levels.
%\newline [\textit{145 words}]

\vspace{0.4cm}
\end{abstract}

\end{onehalfspacing}

\thispagestyle{empty}
\clearpage
\setcounter{page}{1}

%%%
%\clearpage
\section{Introduction}
\label{sec:intro}

The German minimum wage has been a subject of significant interest and debate since its introduction in 2015 at a nominal rate of \euro{8.50}; see \citeA{Baumann2024} for a review. Over time, the minimum wage has undergone gradual increases, eventually reaching \euro{9.82} at the beginning of 2022. However, these adjustments primarily kept pace with rising price levels and overall wage developments in the economy, implying that the minimum wage remained relatively constant in real terms. This tendency changed in the aftermath of the 2021 federal election when a substantial 22 percent increase was implemented in 2022, raising the minimum wage to \euro{10.45} in July and further to \euro{12} in October. This rise to the highest PPP-adjusted minimum wage level within the European Union presents an unprecedented opportunity to investigate the effects of this substantial policy change, which affected about 4.4 million jobs (14 percent of the workforce in the private sector).

The interest in this research stems from the pivotal role of the minimum wage policy in addressing income disparities and enhancing the economic well-being of low-wage workers. Advocates believe that raising the minimum wage can significantly improve the standard of living for vulnerable segments of the workforce, reduce wage and income inequality, and boost worker productivity. Conversely, skeptics express concerns about potential negative repercussions, such as reduced employment opportunities due to increased personnel costs for employers. In this article, we present the first empirical evaluation of the 2022 German minimum wage increase, focusing on its implications for hourly wages, monthly earnings, employment, and working hours. Leveraging data from the newly designed Earnings Survey, which captures around 8 million employee-level observations from a representative sample of establishments each month in 2022, we employ a difference-in-difference-in-differences approach to isolate the causal effects of the minimum wage increase. 

Our findings reveal significant positive impacts on both hourly and monthly wages following the implementation of the minimum wage hike. Low-wage workers experience tangible improvements in their earnings, affirming the intended positive impact of the policy on their economic well-being. Importantly, the employment retention rate (and the regional employment levels) show no significant changes, suggesting that firms effectively adapted to the higher labor costs without resorting to employment reductions.

However, an interesting observation emerges concerning working hours, for which we identify a negative effect in the short run. We find a significant reduction in working hours, in particular for the group of marginal employees (\textit{minijobbers}), which are most severely affected by the minimum wage increase.\footnote{Minijobs are marginal employment relationships which are paid a monthly wage of no more than \euro{450} at the beginning of 2022. These jobs are appealing to both employers and employees because of reduced social security contributions. Additionally, they are exempted from income tax, resulting in a net wage that equals the gross monthly wage. The minijob threshold increased in parallel with the minimum wage from \euro{450} to \euro{520} in October 2022.} This trade-off between wage gains and reduced hours partially offsets the overall positive wage effect when considering monthly earnings, unraveling a nuanced interplay between the minimum wage policy and the labor incomes of employees.

Our findings complement those of the literature on the introduction of the 2015 minimum wage. There is consent in the existing literature regarding positive wage effects of the minimum wage \cite{Baumann2024}, which has been contributing to the decline in wage inequality \cite{Bossler2023}. The most debated issue is a possible (causal) employment reaction after the introduction of the minimum wage. In fact, most studies did not detect any aggregate negative employment effects on the number of jobs \cite{Ahlfeldt2018,Bossler2023,Dustmann2022,Garloff2019}. If anything, the effect was quantified to be relatively small \cite{Bossler2020,Caliendo2018}. However, when distinguishing between regular jobs and minijobs, the evidence suggests a small reduction in the number of minijobs \cite{Caliendo2018,Caliendo2023}, which resulted in about 1 percent of all minijobs transitioning into non-employment \cite{Bossler2024}. The literature is less clear regarding the intensive margin of working hours adjustments. Survey-based studies tend to find a negative effect in the first year but not beyond \cite{Bossler2020,Burauel2020,Ohlert2022}. Based on the more comprehensive information of the Structure of Earnings Survey, the overall hours effect is negligible \cite{Biewen2022,Bossler2024}. However, negative hours responses are observed among those minijobbers who initially worked relatively long hours \cite{Bossler2024}. Overall, this evidence-based evaluation of the minimum wage contributed to the political debate and the decision to increase the real minimum wage substantially for the first time in 2022. 

We add to the literature by presenting the first evaluation of the 2022 minimum wage increase, providing essential policy guidance to policymakers and the Minimum Wage Commission. By uncovering the effects of this substantial wage adjustment on various economic dimensions, our research informs evidence-based decision-making, facilitating the formulation of more targeted and effective labor market policies. While there are several studies on the impact of the \euro{8.50} minimum wage introduction in 2015, their finding cannot necessarily be extrapolated to the 2022 policy which lifted the minimum wage to a much higher level and also took place in a different economic environment. 
While real wages followed a rising trend when the minimum wage was introduced in 2015, they fell in 2022.\footnote{See time series of the \hyperlink{https://www.destatis.de/EN/Themes/Labour/Earnings/Agreed-Earnings-Collective-Bargaining-Coverage/_node.html}{German Statistical Office}.} Similarly, the ifo Business Climate Index was at a much higher level in 2015 than in 2022.\footnote{See time series of the \hyperlink{https://de.tradingeconomics.com/germany/business-confidence}{ifo Business Climate Index}.} While these developments point to a more disadvantageous economic situation, which may lead to a stronger labor demand response to a wage hike, labor market tightness rose until 2022, which may attenuate labor demand responses to rising wages \cite{Beaudry2018,BosslerPopp2024}.

The significance of our research lies in the novelty and richness of the utilized data. Unlike administrative social security data commonly used in prior studies, the Earnings Survey is a monthly panel, providing real-time insights into changes that occur month by month. Additionally, the Earnings Survey data remedies a significant shortcoming inherent in many studies analyzing German administrative employment data, such as \citeA{Bossler2023,Card2013}, and others, namely the absence of information on working hours. The inclusion of working hours data in our analysis provides a comprehensive view of the labor market and offers valuable insights into the interplay between minimum wages and working hours. Moreover, responses to the Earnings Survey are mandatory by German law, ensuring a high level of data completeness and accuracy. This further enhances the reliability of our findings, enabling us to draw robust conclusions from our empirical analysis.

As another contribution of our research, the data enables us to assign employees to narrowly defined treatment groups based on their hourly wages. This precise definition of treatment groups on the individual level allows us to estimate effect heterogeneities at the individual level, such as by gender, education, or employment status. This granular analysis provides valuable insights into how the minimum wage increase affects different subgroups of workers, offering a more nuanced understanding of the policy's impact on wage disparities and income distribution.

The paper proceeds as follows: 
Section~\ref{sec:inst} explains the institutional background. 
Section~\ref{sec:data} describes the Earnings Survey, the data source of this study. 
Section~\ref{sec:method} provides a description of the difference-in-difference-in-differences approach, the methodology used to uncover the effects of the 2022 minimum wage increase. 
Section~\ref{sec:results} presents and discusses the baseline results and several robustness checks, while Section~\ref{sec:heterogeneities} provides further analysis. 
Section~\ref{sec:conclusion} concludes.

\section{Institutional background}
\label{sec:inst}

In Germany, wages were traditionally set and regulated by collective bargaining between industry-specific employer associations and unions.\footnote{In a few industries, employers and unions even agreed on sectoral minimum wages, which existed since the 1990s, but their scope in the low-wage segment was very limited. For empirical evaluations, see e.g., \citeA{Kunaschk2024} for hairdressing and \citeA{Moeller2012} as well as \citeA{fitzenberger2016conceptual} for an overview.} Since employers' voluntary participation in their respective associations' collective bargaining declined over time, the workers' collective bargaining coverage deteriorated \cite{Oberfichtner2019}. This trend has been regarded as one of the reasons for the increasing wage inequality and an increasing share of the working poor \cite{bosch2008low,Dustmann2014}. These developments ultimately led to the introduction of the national minimum wage in 2015, after the 2013 election was decided in favor of political parties that supported the minimum wage \cite{Garloff2019}. 

\begin{figure}[ht!]
\centering
\caption{Development of gross and real minimum wage in Germany}
\label{fig:mw_development}
\includegraphics[width=0.7\textwidth]{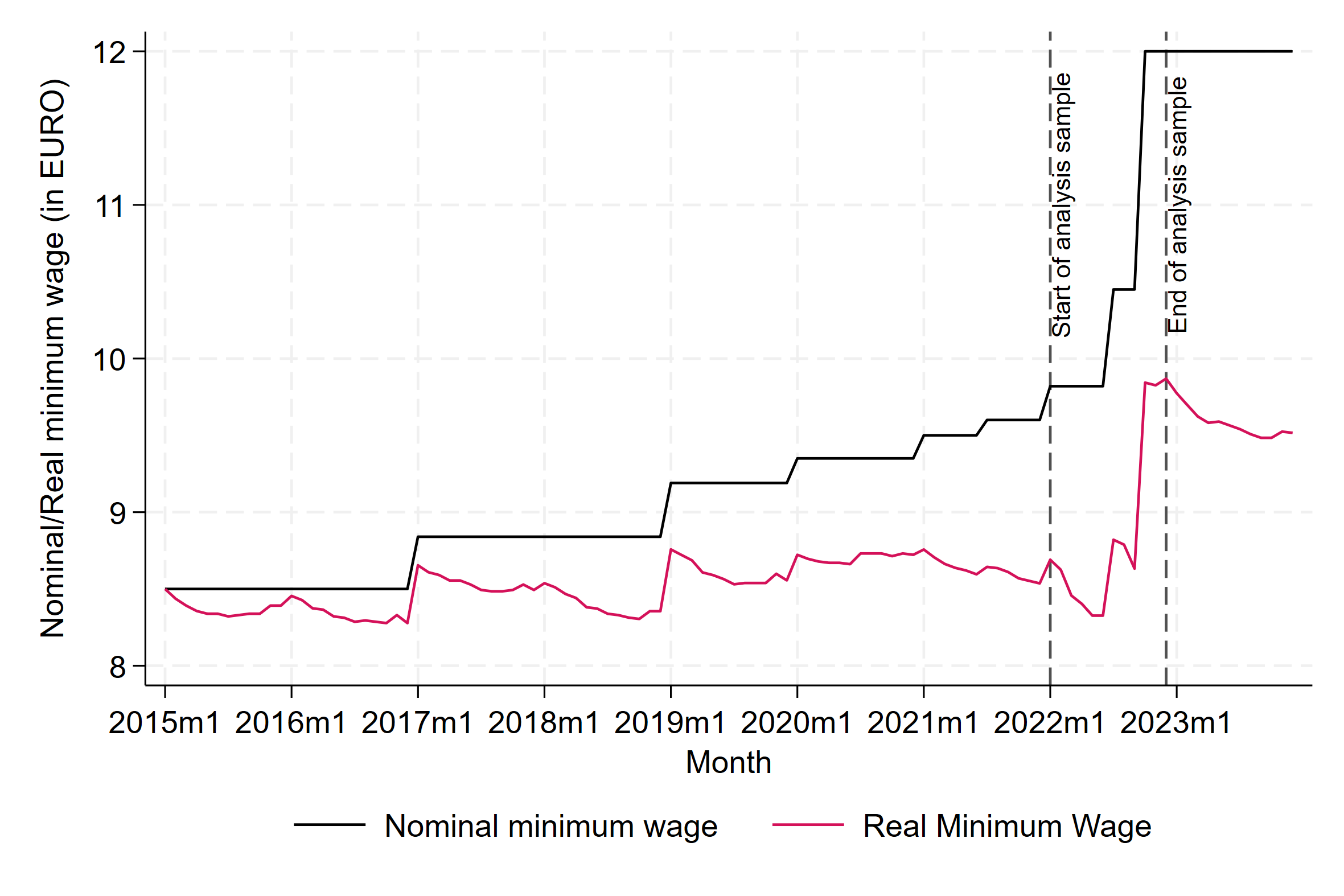}
\begin{justify}
\begin{small}
{\textit{Notes:} Nominal and real minimum wage by month. The real minimum wage is CPI-deflated to \euro-values of January 2015. Labels on the x-axis denote the first month (\textit{m1}) of each year.}
\end{small}
\end{justify}
\end{figure}

The initial level of the minimum wage was politically set at the nominal hourly wage rate of \euro{8.50}. At the same time, a commission was implemented to monitor and evaluate the minimum wage. The German Minimum Wage Commission consists of an equal representation of members from employer associations and unions and it has a mandate to propose adjustments of the minimum wage level every two years, which is then put into law by an ordinance of the government. Up to 2022, this procedure led to modest nominal increases, as depicted by Figure \ref{fig:mw_development}, which mostly followed the average development of collectively bargained wages and the price development \cite{Boerschlein2023}. 

In 2022, for the first time, the government decided to increase the minimum wage by a larger margin without confessing the expertise of the Minimum Wage Commission. This led to a politically motivated minimum wage increase of 22 percent in two steps. In the first step, which was still in accordance with the proposed adjustments of the Minimum Wage Commission, in July 2022, the minimum wage was raised from \euro{9.82} to \euro{10.45}. In the second step, which will be the main focus of our analysis, in October 2022, the minimum wage was raised from \euro{10.45} to \euro{12}. Figure \ref{fig:mw_development} demonstrates that the second increase of 2022 implied the first meaningful hike in real terms, whereas the previous minimum wage adjustments mostly compensated for the devaluation of nominal wages. 

While the 2015 minimum wage introduction has been evaluated in several studies, see for a review \citeA{Baumann2024}, the subsequent smaller nominal increases did not provoke significant economic effects on labor market outcomes \cite{Baumann2024,Boerschlein2022,Caliendo2023}, which corresponds with their negligible relevance in real terms. The 2022 increases, which resulted in the German minimum wage being the highest among all European countries in terms of its PPP-adjusted value (see Appendix Figure~\ref{fig:Europe}), set the stage for the analyses in this article.

\section{Data}
\label{sec:data}

The main data source of our empirical analysis is the monthly Earnings Survey (ES) collected and provided by the Federal Statistical Office of Germany (Destatis). The ES was explicitly designed to improve the accuracy of minimum wage research. In January 2022, the ES replaced the Quarterly Earnings Survey (QES), the four-year Structure of Earnings Survey (SES), and the yearly Voluntary Earnings Survey (VES), combining their characteristics, see \citeA{Finke2023}.

Participation in the ES is mandatory under the legal basis of the Earnings Statistics Act (VerdStatG). The monthly survey sample of the ES that we are using covers approximately 58,000 establishments across all private sectors with at least one employee subject to social insurance.\footnote{Establishments employing only minijobbers, establishments which are just private households, and extraterritorial units (such as embassies) are not included in the survey. Note that imputed employment information for establishments employing only minijobbers is added to the April waves. We do not use this information in our analysis. } The data that we are using excludes the public sector.\footnote{For reporting purposes of Destatis, a sample of staff statistics ("Personalstandsstatistik") collected once a year in June is projected each month to complement the ES with employment information from the public sector (i.e., sectors O and P). The staff statistics provide comprehensive data on all public sector employees but are collected by a different method.} 

While the establishments' personnel departments are the respondents of the ES, the principal purpose of the survey is to collect information at the employee level, capturing information on earnings, working hours, gender, year and month of birth, nationality, level of education, employment group, company tenure, and a 5-digit occupation code ("Taetigkeitsschluessel"). Furthermore, the data include establishment-specific information on the establishment size (measured by the total number of employees), the sector of operation, the extent of collective bargaining coverage, and the location of the establishment. Starting from January 2022, the ES covers all employees from the participating establishments in each month. Unique panel identifiers (personnel numbers and establishment identifiers), in combination with the date of birth, allow for the tracking of individuals over time (if they remain at the same employer). These new and comprehensive data enable a detailed and highly frequent examination of the German labor market in the private sector, and it is perfectly suited to analyze employee-level wages, working hours, monthly labor earnings, and employment stability. 

To ensure data representativeness, the ES provides a weight which is constant across employees in a given establishment and month. The weighting takes into account the sampling matrix, which is based on the federal state, the economic sector, and the size of the establishment, and the weights are constructed by dividing in each cell the number of establishments in the population by the number of establishments in the sample. 
Throughout the paper, we use these weights when computing descriptive statistics and least squares estimates. Our analysis sample covers January to December 2022. We restrict the sample to individuals eligible for the national minimum wage and, therefore, exclude apprentices, interns, and employees below the age of 18. The data include information on the number of paid working hours per month, excluding overtime, with a correction for the number of paid working days per month. We adjust gross monthly earnings following the definition of the Minimum Wage Commission and exclude overtime compensation, premiums for shift, night, Sunday, and/or holiday work, and extra payments.\footnote{The energy price compensation payments in September 2022, which
have been a gross lump sum payment of \euro{300} paid by the employers who were fully compensated by the state,
are not reported in the ES.} Gross hourly wages are obtained from adjusted gross monthly earnings and corrected paid working hours.

\afterpage{
\begin{landscape}

\vspace*{\fill}

\begin{table}[ht!]
\begin{center}
\caption{Descriptive statistics} \label{tab:description}
\resizebox{1.0\linewidth}{!}{
\begin{threeparttable}
\begin{tabular}{l ccc ccc ccc ccc}
\hline \hline
    & Jan.  &   Feb.&  March&   April&   May   &  June &   July  &  Aug.&   Sept.&   Oct.&   Nov.&   Dec. \\[1ex]
\hline
Gross hourly wage                     &  21.76 &   22.07  &  21.54  &  21.89  &  21.76  &  21.80  &  22.00  &  21.60  &  21.97  &  22.56  &  22.97 &   23.40     \\
Log gross hourly wage                 &   2.940&    2.952 &   2.933 &   2.952 &   2.944 &   2.947 &   2.962 &   2.951 &   2.967 &   2.997 &   3.015&    3.030    \\[1ex]
Gross monthly wage                    &  2969.1&  2956.76&  3036.11&  3008.08&  3034.66&  3044.54&  3043.42&  3086.17&  3096.54&  3123.50&  3225.9&   3270.87   \\
Log gross monthly wage                &   7.631&    7.630 &   7.663 &   7.652 &   7.663 &   7.667 &   7.668 &   7.684 &   7.690 &   7.708 &   7.744&    7.753    \\[1ex]
Monthly paid working hours            &  129.43&   127.17 &  134.31 &  129.88 &  132.62 &  132.66 &  130.67 &  134.90 &  133.01 &  131.13 &  133.30&   132.52    \\[1ex]
Minijobber                            &   0.131&    0.132 &   0.131 &   0.134 &   0.135 &   0.134 &   0.135 &   0.134 &   0.135 &   0.135 &   0.133&    0.132     \\[1ex]
Employed in month $m+3$ in same plant &  0.853 &   0.867  &  0.867  &  0.884  &  0.879  &  0.861  &  0.880  &  0.874  &  0.869  &         &        &              \\[1ex]
\hline
\multicolumn{13}{l}{Fractions of workers by wage bins:} \\
\quad < 7.50         &   0.012 &  0.011 &  0.011&   0.006 &  0.006 &  0.005 &  0.005 &  0.005 &  0.005 &  0.004 &  0.004 &  0.004  \\[1ex]
\quad 7.50 - 9.81    &   0.025 &  0.020 &  0.019&   0.019 &  0.017 &  0.017 &  0.011 &  0.011 &  0.010 &  0.008 &  0.008 &  0.007  \\[1ex]
\quad 9.82 - 10.44   &   0.056 &  0.057 &  0.056&   0.057 &  0.058 &  0.056 &  0.023 &  0.019 &  0.017 &  0.006 &  0.005 &  0.004  \\[1ex]
\rowcolor{blue!10}
\quad 10.45 - 11.99  &   0.089 &  0.087 &  0.087&   0.086 &  0.086 &  0.085 &  0.115 &  0.115 &  0.110 &  0.036 &  0.031 &  0.027  \\[1ex]
\quad 12.00 - 12.99  &   0.059 &  0.059 &  0.061&   0.062 &  0.063 &  0.064 &  0.066 &  0.068 &  0.064 &  0.115 &  0.108 &  0.108  \\[1ex]
\quad 13.00 - 14.99  &   0.103 &  0.103 &  0.108&   0.106 &  0.107 &  0.107 &  0.107 &  0.111 &  0.108 &  0.132 &  0.129 &  0.128  \\[1ex]
\quad 15.00 - 20.99  &   0.284 &  0.281 &  0.294&   0.288 &  0.295 &  0.294 &  0.292 &  0.301 &  0.303 &  0.302 &  0.300 &  0.293  \\[1ex]
\quad $\geq$ 21.00   &   0.372 &  0.383 &  0.364&   0.377 &  0.368 &  0.371 &  0.381 &  0.370 &  0.383 &  0.397 &  0.417 &  0.430  \\[1ex]
\hline
Observations & 
8,047,011 &	7,826,586 &	7,710,347 &	8,184,488 &	8,197,444 &	8,103,563 &	8,323,683 &	8,292,750 &	8,195,775 &	8,377,516 &	8,311,185 &	8,160,975 
\\[0.5ex]
Employees (weighted) &
30,650,459 &	30,710,962 &	30,763,247 &	30,921,648 &	31,083,105 &	31,201,096 &	31,104,984 &	31,090,451 &	31,080,168 &	30,889,643 &	30,795,487 &	30,450,172 
\\[1ex]
\hline \hline
\end{tabular}
\begin{small}
\textit{Notes:} Average values by month are based on the weighted data using the simple inverse probability weighting factor.\\
\textit{Source:} Earnings Survey, monthly panel January-December 2022, public sector excluded. 
\end{small}
\end{threeparttable}
}
\end{center}
\end{table}

\vspace*{\fill}

\end{landscape}
}

The data set includes nearly 100 million employee-level observations, which are evenly distributed across months, with each month accounting for approximately 7.9 percent to 8.6 percent of the observations. Hence, the data comprises about 8 million workers employed in about 50,000 establishments each month.\footnote{While the gross sample includes 58,000 private sector establishments, about 3,300 did not respond because they closed or can no longer be found and 4,600 did not respond although being required to respond by law, corresponding to a \textit{genuine} non-response rate of 8.5 percent (ES data of April).} The upper panel of Table~\ref{tab:description} shows the 2022 monthly weighted averages of employee-level labor market outcomes. Average hourly wages are initially at \euro{21.8} and increase on average by about 9 percent from the beginning to the end of the year. Two-thirds of this increase takes place after the minimum wage hike in October. Average monthly wages are just below \euro{3,000} in January 2022 and increase by 12 percent until December. Again, a significant increase (of 6 percent) is obtained in the last three months, i.e., after the minimum wage increase in October. Regarding paid working hours, there is no clear trend visible. Rather, hours fluctuate according to the working days in a particular month, e.g., they are larger in March (23 working days) and smaller in February (20 working days). About 1 out of 8 employment relationships in the private sector is a minijob. The probability of being employed in the same plant three months later varies between 85 and 88 percent. Hence, at the aggregate level, we do not observe reduced employment retention after the minimum wage increases, which would manifest, e.g., in a lower observed rate in July (to be employed in October). Of course, this may change once we compare narrowly defined treatment and control groups.

In further analyses presented in Section \ref{sec:heterogeneities}, we also analyze total employment at the region level. For the regional analysis, we exploit county-level administrative employment data from the Federal Employment Agency, where counties are administrative districts ("Kreise"), of which there are 400 in Germany. The data include the total number of all employees, allowing us to distinguish between regular jobs and minijobs. The county data on the number of jobs is monthly accurate as they are counted based on mandatory reports to the social insurance of all employers in Germany on each of their employees.

\section{Method}
\label{sec:method}

The precise and high-frequency information on hourly wages at the employee level, which has not existed before but is now available in the new ES data, allows us to compare narrowly defined groups of employees based on their hourly wage before the minimum wage increases. This allows us to define two treatment groups of employees, which can be compared to various narrowly defined control groups.

\paragraph{Treatment group definition.}
 
Our main focus is on employees who are only affected by the second minimum wage hike during 2022, namely by the increase to \euro{12} in October 2022, while they are paid at or above \euro{10.45}, which is the minimum wage that came in force in July.
Hence, our treatment group of main interest is defined as follows:
 \begin{itemize}
\item \euro{12.00}-MW-Treatment: $ {\mbox {\euro}}10.45 \leq w_{i,m} \leq {\mbox {\euro}}11.99$
\end{itemize}
where $i$ indexes workers and $m$ months. The group composition may have changed in the course of the year due to inflows and outflows of the labor market and due to employees changing the wage groups. We address the latter by a time-varying group assignment. 
As can be seen from the lower panel of Table~\ref{tab:description}, as of September 2022, the \euro{12.00}-MW-Treatment group comprises 11 percent of the employees. Further, the descriptions show that after the increase of the minimum wage to \euro{12.00} in October, the size of the group drops to 3.6 percent and 2.7 percent until December, demonstrating the effectiveness of the minimum wage increase to lift these employees up along the hourly wage distribution.

In later analysis (reported in Section~\ref{sec:heterogeneities}), we also examine a second (smaller) treatment group, namely those paid below the July minimum wage level of \euro{11.45} but above the minimum wage of \euro{9.82} which was still effective at the beginning of the year. Hence, this second treatment group is defined as follows:
 \begin{itemize}
\item \euro{10.45}-MW-Treatment: $ {\mbox {\euro}} 9.82 \leq w_{i,m} \leq {\mbox {\euro}}10.44$
\end{itemize}
The \euro{10.45}-MW-treatment group captures between 5.6 and 5.8 percent from January to June (see Table~\ref{tab:description}), indicating that almost 6 percent of the workforce is directly affected by the first minimum wage increase. Correspondingly, the share of this group sharply drops to 2 percent in July. Moreover, its size further shrinks to 0.5 percent in October, indicating a lagged effect experienced by this group after the more prominent minimum wage increase to \euro{12} in October.

\paragraph{Control group definition.}
The data also allow us to distinguish between various control groups of employees. We can compare the treated employees with those paid only slightly above the \euro{12}-minimum wage. This control group allows for close comparison with the treated employees, but it bears the risk that the untreated are themselves affected by the minimum wage through wage spillovers. Therefore, we also compare the treated employees with groups of workers further up in the hourly wage distribution, which reduces the risk that the control group is itself affected. Altogether, we look at the following three control groups. 
\begin{itemize}
\item   Control group 1: $ {\mbox {\euro}}12.00   \leq w_{i,m}  \leq  {\mbox {\euro}}12.99 $
\item   Control group 2: $ {\mbox {\euro}}13.00   \leq w_{i,m}  \leq  {\mbox {\euro}}14.99 $
\item   Control group 3: $ {\mbox {\euro}}15.00   \leq w_{i,m}  \leq  {\mbox {\euro}}20.99 $
\end{itemize}
It is evident from Table~\ref{tab:description} that the size of the three control groups is relatively stable between January and September, i.e., before the minimum wage hike in October. This is a precondition to serve as appropriate counterfactuals. 

\paragraph{Further groups.}
Those worker-month observations which do not fall into either of the two treatment groups or the three control groups defined above are categorized into three further bins.
These groups are less relevant for our research question, for which reason they will not be part of our graphical inspections throughout Sections~\ref{sec:results} and \ref{sec:heterogeneities} (while still being included in the accompanying regressions). In a first group, we define high-wage workers as those earning a wage of at least \euro{21}. While the median employee belongs to this group (see Table~\ref{tab:description}), it is obviously not suitable as a control group in minimum wage analyses. Two additional groups capture workers with a wage below \euro{9.82}, i.e, below the minimum wage level effective as of the beginning of 2022. Hence, irrespective of the month these observations imply that workers are paid non-compliant such that it is not clear whether these workers would benefit from a further increase of the minimum wage (i.e., they are not an interesting treatment group).\footnote{Note that wages below the minimum wage can also be due to measurement error if working hours are inaccurately reported or if individuals are in fact exempted from the minimum wage.} We distinguish between workers paid just below the minimum wage being in force at the beginning of the year (i.e, who are paid between \euro{7.50} and \euro{9.81}) and those workers who are paid much below the initial minimum wage (i.e., who are paid below \euro{7.50}). According to Table~\ref{tab:description}, in January 2022, these two groups at the very bottom of the wage distribution comprise only 3.3 percent and the fraction continuously decreases throughout the year.

\paragraph{Regression specification.}

In our main empirical analysis, we estimate effects on four individual-level labor market outcomes: hourly wages, monthly wages, working hours, and employment retention. We specify the first three of these outcomes as the difference in the logarithm between month $m$ and month $m+3$, such that they measure (approximately) the respective percentage change over three months. Since our panel data consists of twelve monthly waves in 2022, we have nine months (January to September) for which we can calculate the individual growth rates over three months. We apply a flexible difference-in-differences specification, which is fully saturated in nine months dummies $M_{k,m}$ and eight worker group dummies $G_{g(im),m}$:
\begin{eqnarray}\label{eq:didid}
\ln y_{i,m+3} - \ln y_{i,m} = \alpha_{1} + \sum_{k = 2}^{9} \alpha_{k} M_{k,m}  + \sum_{g=1, g\neq5}^{8} \,\, \sum_{k=1}^{9}  \beta_{g,k}  G_{g(im),m} M_{k,m}  + u_{i,m}  \\[2ex]
\qquad \qquad m = 1, \ldots, 9 \text{ months}   \nonumber
\end{eqnarray}
where $y_{i,m}$ denotes either hourly wages, monthly wages, or working hours of worker~$i$ at month~$m$.\footnote{Obviously, various reparameterizations of the fully interacted model are possible, each containing 72 parameters. Estimating equation~(\ref{eq:didid}) yields the treatment effect as follows. Group 4 is the 12.00-MW-Treatment Group ($10.45 - 11.99$), and group 5 is the first control group beyond the minimum wage ($12 - 12.99$). Hence, $\beta_{4,7}$ (respectively $\beta_{4,8}$ or $\beta_{4,9}$) minus $\beta_{4,6}$ produces the treatment effect. Predictions of the dependent variables by group and month can be obtained from linear combinations of the estimates of $\alpha_{m}$ and $\beta_{g,m}$.} By construction, growth rates can only be calculated if individuals are observed in $m$ and $m+3$ in the same establishment, ruling out compositional changes influencing the growth rates. 

The construction of the dependent variable as a log difference closely follows the approach presented in \citeA{Dustmann2022}, who use a yearly forward-looking difference instead of within-year three-month differences. In fact, the estimation yields a difference-in-difference-in-difference specification in log wages since another difference is added through the construction of the dependent variable. The first advantage of the specification over a naive difference-in-difference estimation is that it cancels out differences in wage growth between wage groups that occur irrespective of the minimum wage. This is important since the prevalence of differential wage growth is well documented and often referred to as mean-reversion, i.e., low-wage workers' wages would have grown stronger even in the absence of a minimum wage.\footnote{See, e.g., \citeA{Derenoncourt2022} for an example in which mean-reversion has been influential for the effects of minimum wages. But mean-reversion is prevalent in many other settings, such as in the catch-up of earnings after job loss \cite{Jacobson1993}.} Note, however, that this formulation of the model also causes a crucial change to the identifying assumption, which requires parallel trends in the three-month growth rates instead of parallel trends in the levels. The second advantage of the empirical specification is that it allows an analysis of very short-term jobs which have been existing only for a few months. This is an advantage over the conventional approach of fixing the bite according to the wage in one specific month, which requires that the jobs already existed in that particular month, leading to an increasingly selective sample over time. Nevertheless, we provide a robustness check, which fixes the bite as of June ensuring that the effects of the minimum wage are not influenced by an endogenous change of the treatment definition due to the smaller minimum wage increase of July. 

To examine whether the minimum wage increase has affected employment stability, we estimate the following linear probability model:
\begin{eqnarray}\label{eq:did_Ret}
E_{i,m+3}  =  \gamma_{1} + \sum_{k=2}^{9}\gamma_{k}  M_{k,m}   + \sum_{g=1, g\neq5}^{8} \,\, \sum_{k=1}^{9} \delta_{g,k} G_{g(im),m} M_{m,k}  + \nu_{i,m} 
\\[2ex]
\qquad \qquad m = 1, \ldots, 9 \text{ months} \nonumber
\end{eqnarray}
where $E_{i,m+3}$ is a dummy variable equal to one if a worker~$i$ is observed to be employed in the same establishment also three months ($m+3$) afterwards. We estimate Equations~(\ref{eq:didid}) and~(\ref{eq:did_Ret}) by weighted least squares using the supplied sampling weight to ensure the representativeness of our results. Standard errors are clustered at the individual level.\footnote{We were restricted to using the software package SAS. Since an option for clustered standard errors is unavailable, we computed them ourselves.} We will inspect the estimated effect patterns graphically by plotting the four predicted outcome variables for each month-group combination of the treatment group and the control groups.

\section{Results} \label{sec:results}

\begin{figure}[ht!]
\centering
\caption{Baseline hourly and monthly wage effect} \label{fig:12_base_wage}
\includegraphics[width=0.7\textwidth]{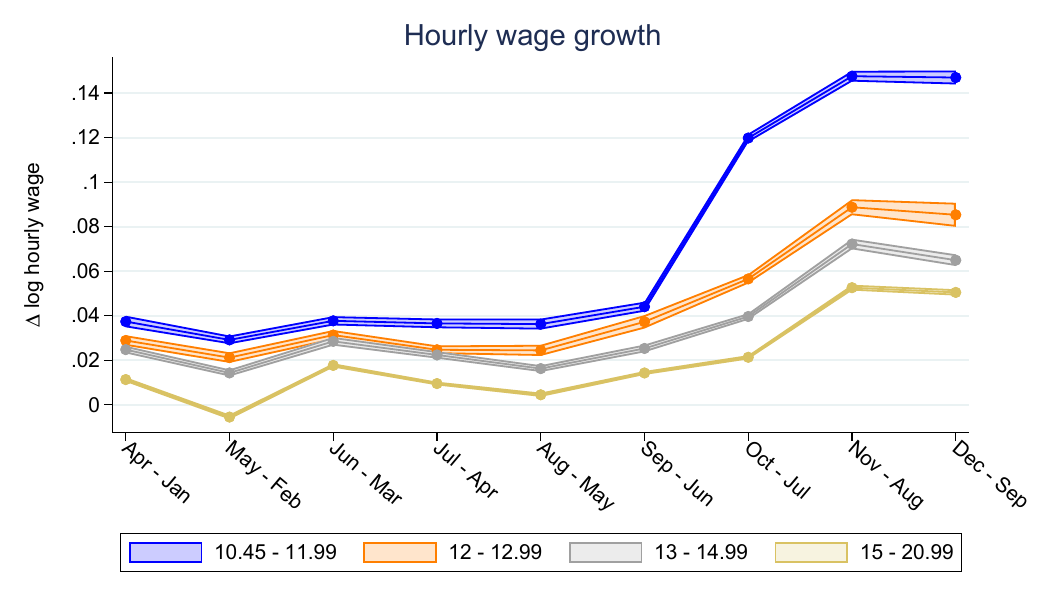}
\includegraphics[width=0.7\textwidth]{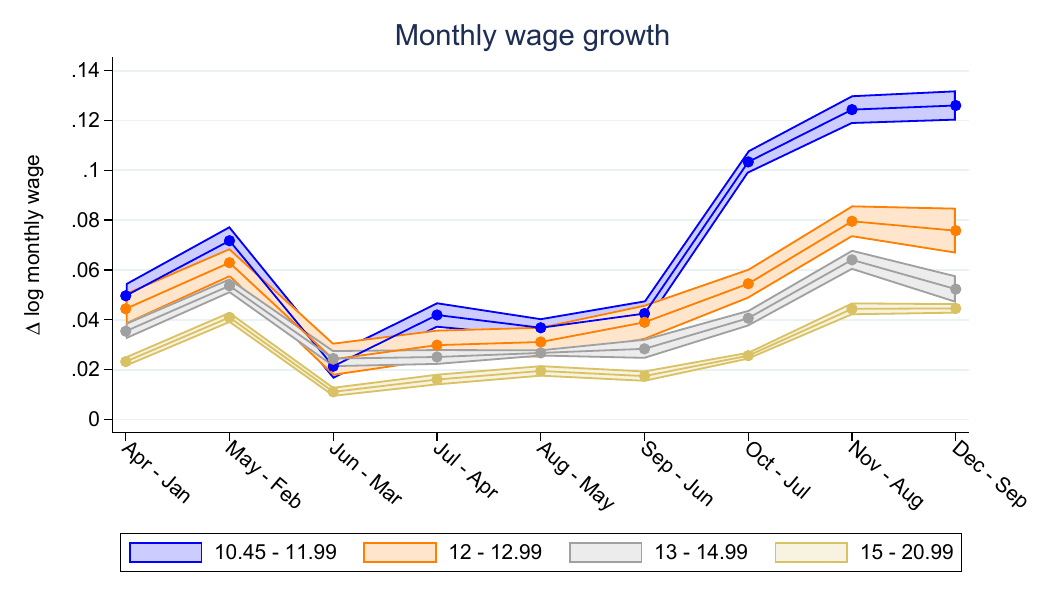}
\begin{justify}
\begin{small}{\textit{Notes:} The upper part presents the development of hourly wage growth and the lower part presents the development of monthly wage growth. Each point is a prediction using the estimates of equation~(\ref{eq:didid}) for the respective group $g$ and month $m$, i.e., each point is an estimate of $E[\ln y_{g,m+3} - \ln y_{g,m}]$. Shaded areas show the 95\%-confidence intervals with standard errors clustered at the individual level. \\
\textit{Source:} Earnings Survey, monthly panel January-December 2022, public sector excluded. }
\end{small}
\end{justify}
\end{figure}

\paragraph{Effects of the \euro{12} minimum wage.} We start by analyzing the wage effects of the major minimum wage increase from \euro{10.45} to \euro{12} taking place in October 2022. Using estimates of equation~\ref{eq:didid}, Figure~\ref{fig:12_base_wage} illustrates predictions of hourly wage growth (in the upper part) and of monthly wage growth (in the lower part). The groups included in the graph are the \euro{12}-treatment group (${\mbox {\euro}}10.45 \leq w_{i,m} \leq {\mbox {\euro}}11.99$), which is affected because these workers' wages needed to be lifted to the new minimum wage level, and three control groups of workers which are all located above the new minimum wage of \euro{12}. The full table of regression estimates, including all groups of workers, is presented in Appendix Table~\ref{tab:app_baselinereg}.

The comparison of the treatment group (blue line) with the three control groups shows the treatment effect.\footnote{More precisely, the treatment effect is the change in the gap between the blue line and the other lines before and after the treatment takes place. Since the treatment (i.e., the minimum wage hike) takes place in October, the first growth rate after treatment is \textit{Oct-Jul} while the last growth rate before treatment is \textit{Sep-Jun}.} For hourly wages, the treatment group's wages rose sharply in October, exactly when the minimum wage was raised. Compared with the control groups, the hourly wage growth effect can be quantified as 5.7 percentage points on average (with a standard error of 0.002), which also corresponds to an increase in wage levels of 5.7 percent. The effect remains stable over the three post-treatment months. Similarly, the lower part of Figure~\ref{fig:12_base_wage} also shows a meaningful treatment effect in monthly wage growth since a gap of 4.5 percentage points opens up between the treatment and the control groups in October (with a standard errorof 0.005). The effect size for monthly wages is slightly smaller than for hourly wages, which suggests a small negative hours effect of the \euro{12}-minimum wage hike. 

In addition, Figure~\ref{fig:12_base_wage} shows that the three-month growth rates evolved parallel for the four groups before the treatment came into force in October, giving credence to the parallel trends assumption. Moreover, Figure~\ref{fig:12_base_wage} also depicts (time-constant) differences in wage growth between the four groups of interest. This difference in growth rates suggests that the parallel trends-assumption in wage levels would likely be violated. The higher growth rates of the lower-paid groups indicate that they catch up in levels, likely due to mean-reversion. In sum, the assessment of trends favors our specification in growth rates over an estimation in levels.

Concerning the remaining groups of workers included in specification~(\ref{eq:didid}), note that we analyze and interpret the group that is affected by the \euro{10.45}-minimum wage increase separately in Section~\ref{sec:heterogeneities}. Further, Appendix Table~\ref{tab:app_baselinereg}) documents that workers who were even paid below the initial minimum wage of \euro{9.82}, i.e., who were paid non-compliant, do not benefit from the minimum wage hikes if paid far below the minimum wage ($ w_{i,m} \leq {\mbox {\euro}} 7.49 $), but they tend to also benefit if paid only slightly below minimum wage (${\mbox {\euro}} 7.50 \leq w_{i,m} \leq {\mbox {\euro}}9.81$).

\begin{figure}[ht!]
\centering
\caption{Baseline working hours and employment retention effect} \label{fig:12_base_employment}
\includegraphics[width=0.7\textwidth]{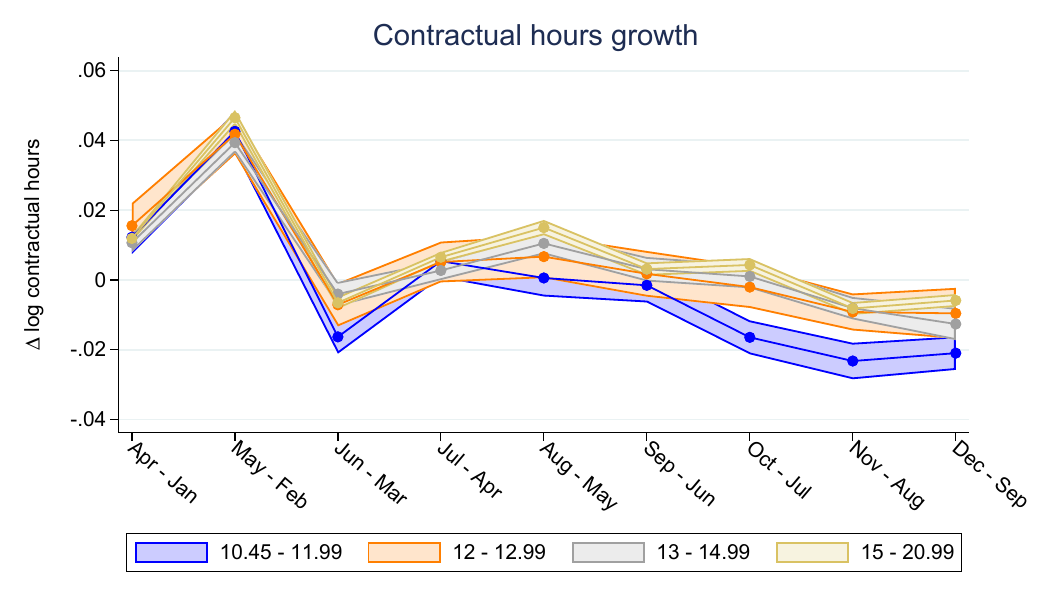}
\includegraphics[width=0.7\textwidth]{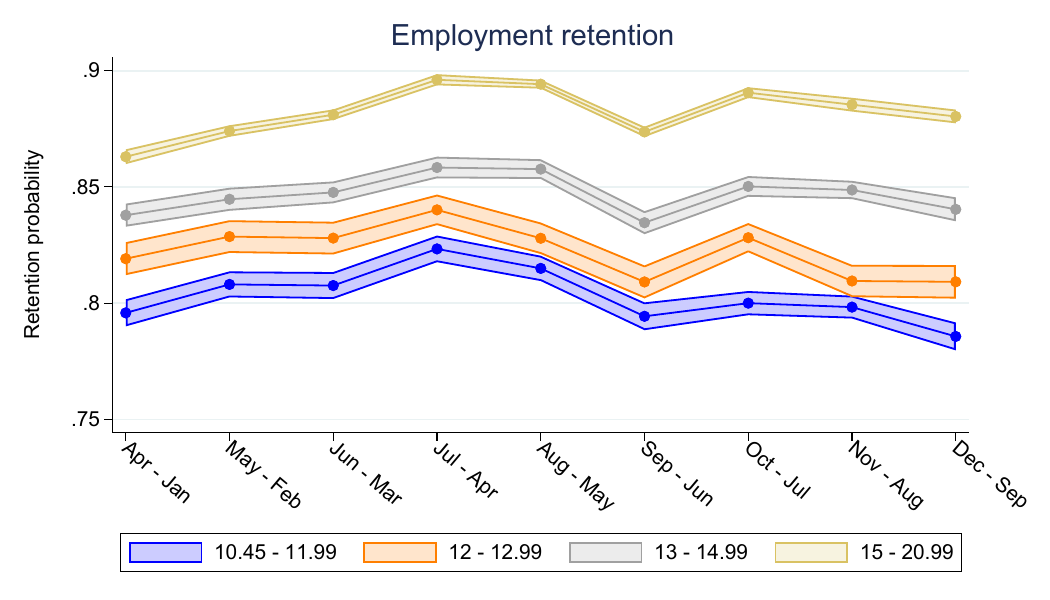}
\begin{justify}
\begin{small}{\textit{Notes:} The upper part presents the development of monthly working hours growth and the lower part presents the development of employment retention. For monthly working hours, each point is a prediction using the estimates of equation~(\ref{eq:didid}) for the respective group $g$ and month $m$, i.e., each point is an estimate of $E[\ln y_{g,m+3} - \ln y_{g,m}]$. For employment retention, each point is a prediction using estimates of equation~(\ref{eq:did_Ret}) for the respective group $g$ and month $m$, i.e., each point is an estimate of $E[\ln E_{g,m+3}]$. Shaded areas show the 95\%-confidence intervals with standard errors clustered at the individual level.\\
\textit{Source:} Earnings Survey, monthly panel January-December 2022, public sector excluded.}
\end{small}
\end{justify}
\end{figure}

Figure~\ref{fig:12_base_employment} illustrates the effects of the \euro{12} minimum wage hike on the intensive and the extensive margins of employment, i.e., the figure displays adjustments in monthly working hours (upper part) and in employment retention (lower part). While working hours follow a somewhat peculiar pattern in the first half of the year, this development is explained by a changing number of working days across these months. Most importantly, this change in working hours affects the development of all groups in parallel. However, in October, the \euro{12}-treatment group slightly deviates from the control groups, opening up a negative monthly working hours effect of 1.1 percent (s.e. 0.005). This effect mirrors the difference between the effects on monthly wages and hourly wages, where the former is observed to be one percent smaller than the latter. The lower part of Figure~\ref{fig:12_base_employment} displays the job retention effect, and it seems that a small gap between treatment and control groups again opens up in October. However, this development solely occurs due to changes in the control groups rather than reduced job stability in the treatment group. Moreover, the gap closes back in the two following months. Hence, we argue that the minimum wage increase has no robust impact on the job retention of affected workers. 

In sum, we have documented an hourly wage effect of $+5.7\%$, a monthly wage effect of $+4.5\%$, a working hours effect of $-1.1\%$, and a job retention effect that is negligible and not robust across the three months after the increase. From these effects of the \euro{12} minimum wage hike, we can calculate a labor demand elasticity, which relates the employment effects to the minimum wage-induced hourly wage increase. The calculation of labor demand elasticities (as implied by the empirically realized employment changes) allows for a comparison of estimates across countries and with different settings and sizes of the minimum wage \cite{Dube2019,Dube2024}. Given our empirical effects on wages as well as on job retention and working hours from the same exogenous minimum wage increase, we arrive at a negative employment elasticity of $-0.196$:\footnote{We calculate a standard error of this employment elasticity using the delta method while assuming the absence of covariance between numerator and denominator, see \citeA{Dube2024}. The resulting standard error of the elasticity is 0.089, which makes the elasticity significantly different from zero, given a statistical error probability of 5 percent.} 

\vspace{-0.7cm}
\begin{align}
\eta_{L,w}  \simeq  \frac{\partial \ln  L}{\partial \ln w} & = \frac{\partial  \ln ( \textit{Jobs} \ast   \textit{Hours worked})}{\partial \ln w}  \nonumber \\[2ex]
                 & =  \dfrac{\frac{\partial \ln \textit{Jobs}}{\partial MW} + \frac{\partial \ln \textit{Hours worked}}{\partial MW}}
                 {\frac{\partial \ln \textit{w}}{\partial MW}}  
                 =  \frac{0 + (-0.0111)}{0.0565} = - 0.196 \label{eq:ela_estimate}
\end{align}

This elasticity estimate corresponds with own-wage elasticities across U.S. minimum wage studies reported and analyzed in a meta-study by \citeA{Dube2019}. Regardless of the group of study (e.g., teenagers or all employees), the measurement of employment (e.g., volume or heads), and the minimum wage variation of analysis, the median elasticity is reported to be $-0.196$ across various studies and states. Given that elasticities that are less negative than $-0.4$ are interpreted to be rather small \cite{Dube2019}, we would also interpret our observed hours effect as rather small.\footnote{Note that an adverse hours effect may be interpreted as less detrimental than an employment effect in heads because people may prefer lower working time (which we do not observe information on) and are compensated by leisure. In addition, according to our results the reduction in working hours is more than compensated by the rise in hourly wages.}

Referring to the literature on the German minimum wage introduction, our obtained elasticity is again in the ballpark of reported estimates (as summarized in \citeA{Popp2021}). While some studies do not find any significant employment effect \cite{Ahlfeldt2018,Dustmann2022,Garloff2019}, the employment response is only slightly negative in other studies \cite{Caliendo2018} and points to elasticities between $-0.1$ (\citeA{Bossler2024} for the group of minijobbers) and $-0.3$ \cite{Bossler2020}, where the latter estimate is identified for firm-level variation which may be overestimated due to reallocation of minimum wage workers across firms \cite{Dustmann2022}. While information on working hours was scarce at the time of the minimum wage introduction, the existing studies point to none \cite{Biewen2022} or (for specific groups) only slightly negative estimates \cite{Bossler2024}. With the minimum wage increase to \euro{12}, for the first time, we are able to combine effects from various margins of employment adjustment and combine it with a precise estimate of the wage effect, arriving at an employment elasticity of $-0.196$.

\paragraph{Robustness checks.}

We conduct various checks to test the robustness of the results. 

First, we fix the group assignment of individuals across all months using the wage from June and estimate the model in 4-month differences. This excludes the possibility that the effects might be partly driven by the mobility of workers across wage groups (i.e., composition effects). In particular, fixed group assignment rules out that the estimated effect of the \euro{12 } minimum wage hike could be biased from the intermediate minimum wage increase to \euro{10.45} in July.\footnote{When comparing \textit{Oct-Jul} with \textit{Sep-Jun} a difference can occur because of a change in October (when the minimum wage hikes) or in July. If the effect is due to the base period, i.e., a change in July, the treatment effect would be due to a changing group composition (maybe due to the intermediate minimum wage increase) rather than a genuine treatment effect. However, this concern seems not to be a problem in our case since the difference \textit{Oct-Jul} is robust when we instead fix the groups to wages as of June and further estimate the effect in 4-month differences such that the composition in the base month is fixed over time.} However, it comes at the disadvantage that the sample becomes increasingly selective the further away an observation is from June since the fixing requires that the respective individuals in the sample have already been employed at the same firm in June. The results in Appendix~\ref{app:fixed_groups} demonstrate this selectivity across months when looking at the development of employment retention, which is economically difficult to interpret. Most interestingly, the hourly wage and working hours effects are decreasing over the three post-treatment months. There are two explanations for the short-run decline in effect sizes. Firstly, we believe that the selection effect plays a crucial role since short-run jobs (i.e., jobs that started after June) drop out of the sample.\footnote{When we further restrict the sample to even more stable jobs, i.e., restricting the data to a fully balanced sample, the effect size further deteriorates. This underlines the importance of an identification strategy that includes such short-run jobs.} Secondly, it could also be a genuine deterioration of the minimum wage effects over time due to inflation-driven catch-up effects in wages of employees in higher wage groups who are not directly affected by the minimum wage. Nevertheless, the robust but very short-run decline in working hours demonstrates the importance of analyzing timely adjustments to the minimum wage from high-frequency data. 

Second, we add control variables for gender, age, education, foreign, job task, part-time, minijobber, job tenure, industry, bargaining agreement, company size, and federal state. Since the treatment effect estimates are identified from changes over the months of 2022, they should not be driven by rather time-constant covariates. For this reason, we have not included covariates in our baseline regressions. Nevertheless, the treatment effect could be influenced by a changing employee composition of groups over time. Regarding the working hours effect, changing gender and part-time composition could be a particular concern. Note that all control variables are defined and included in the initial month $m$ so that we can rule out that the control variables are endogenously affected by the minimum wage itself. Compared with our baseline, the respective results presented in Appendix~\ref{app:robustness_cov} are virtually unchanged. 

Third, we examine whether the effects are heterogeneous across observable characteristics and conduct separate analyses by education (3 groups), tenure (4 groups), and establishment size (4 groups). The results are presented in Appendix~\ref{app:robustness_education}---\ref{app:robustness_size}. However, we do not find meaningful differences, suggesting that the baseline effects are fairly stable even across subgroups of minimum wage workers. 

Fourth, we add worker-level fixed effects to our regression specifications. Again, we aim to control for a potentially changing composition of the groups of interest over time. In this respect, the fixed effects estimation ensures that the effects are only identified from changes in the growth rates over time within individuals. The results are very similar to our baseline (see Appendix~\ref{app:robustness_FE}). 

Fifth, we alter the definition of our outcome variables from three-month differences to one-month differences. This leaves us with only one difference overlapping with the minimum wage increase on October 1st, which is the $October - September$ difference. However, the alternative definition of the outcome variable narrows down the treatment effect to a group defined to be treated in September, i.e., closely before the minimum wage was increased. While this alternative specification bears the risk that the treatment effect attenuates due to anticipation effects, we still observe a robust negative impact on the one-month growth rates of monthly working hours while noting that the hourly and monthly wage effects only slightly attenuate (see Appendix ~\ref{app:robustness_1month}).

\section{Additional results}\label{sec:heterogeneities}

\paragraph{Effects on minijobs vs. regular social security jobs.}

In the literature on the German minimum wage, an important distinction has emerged concerning the difference between regular jobs and minijobs. Regular jobs are characterized by regular contributions to social insurance, which are equally shared between employers and employees. By contrast, minijobs' social security contributions are subsidized, as the respective employees receive the gross wage also as their net wage, while employers benefit from unbureaucratic registration. Most importantly, the distinction between regular employees and minijobs is purely based on the monthly salary. Up to the threshold of \euro{450}, the job is a minijob, and above \euro{450}, the respective job is a regular social security job. 

While most of the literature on minimum wages in the U.S. is concentrated on teenage employment (e.g., \citeA{Allegretto2011} or \citeA{Neumark2022}), as these are heavily affected by the minimum wage, teenagers in Germany rarely drop out of the educational system to take up a job for a living, and moreover, in Germany teenagers are exempted from the minimum wage. Instead, minijobs have been identified as a particularly relevant group for the minimum wage. For the latest minimum wage increase to \euro{12}, according to our data, about 40 percent of the minijobs are directly affected.\footnote{Among the treated workers, 50 percent are in a minijob, while minijobs make up only 13 percent of all jobs.} In the studies that evaluated the German minimum wage introduction, it was debated from the beginning whether it destructed minijobs \cite{Caliendo2018,Caliendo2023}. Despite their severe affectedness, minijobbers' gains in monthly wages were smaller if their monthly salary was already right at the \euro{450}-threshold, plausibly because employers or employees wanted to preserve the advantages of the minijobs \cite{Bossler2023}. Consequently, some of these minijobs experienced decreasing hours if they were already paid exactly \euro{450} \cite{Bossler2024}. Moreover, \citeA{Bossler2024} show that about one percent of all minijobbers transitioned into non-employment, notwithstanding that one percent was also upgraded to regular jobs.

\begin{figure}[ht!]
\centering
\caption{Hourly and monthly wage effect for regular employees and minijobbers}\label{fig:12_minijob_wage}
\includegraphics[width=0.99\textwidth]{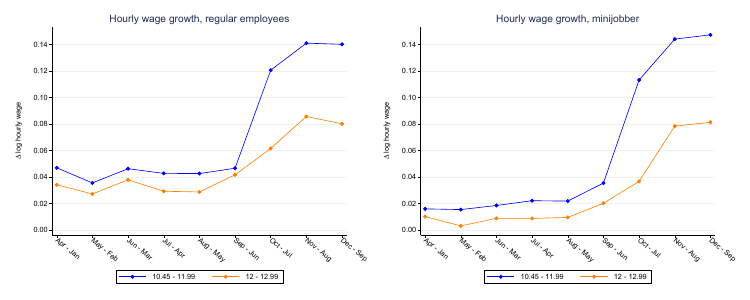}
\includegraphics[width=0.99\textwidth]{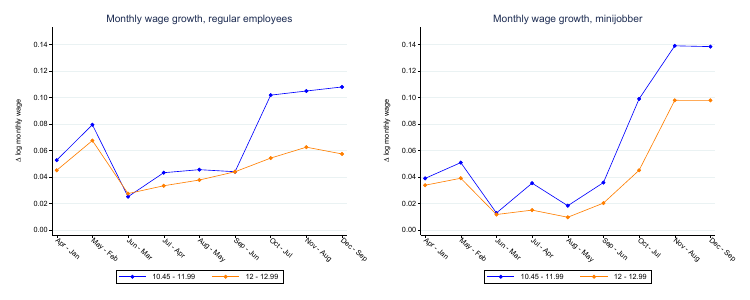}
\begin{justify}
\begin{small}{\textit{Notes:} Separate estimations and predictions for regular social security employees (on the left) and minijobbers (on the right). The upper part presents the development of hourly wage growth and the lower part presents the development of monthly wage growth. Each point is a prediction using the estimates of equation~(\ref{eq:didid}) for the respective group $g$ and month $m$, i.e., each point is an estimate of $E[\ln y_{g,m+3} - \ln y_{g,m}]$. \\
\textit{Source:} Earnings Survey, monthly panel January-December 2022, public sector excluded.\\}
\end{small}
\end{justify}
\end{figure}

\begin{figure}[ht!]
\centering
\caption{Working hours and employment retention effect for regular employees and minijobbers}\label{fig:12_minijob_employment}
\includegraphics[width=0.99\textwidth]{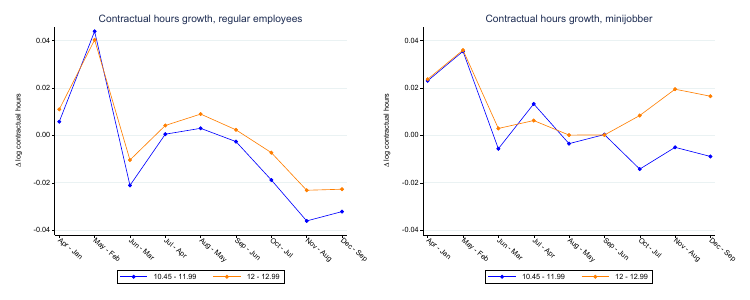}
\includegraphics[width=0.99\textwidth]{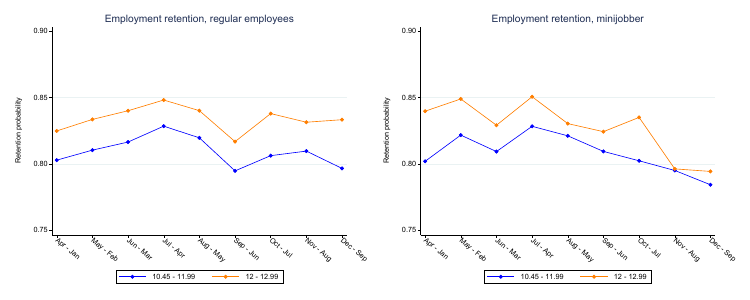}
\begin{justify}
\begin{small} {\textit{Notes:} Separate estimations and predictions for regular social security employees (on the left) and minijobbers (on the right). The upper part presents the development of monthly working hours growth and the lower part presents the development of employment retention. For monthly working hours, each point is a prediction using the estimates of equation~(\ref{eq:didid}) for the respective group $g$ and month $m$, i.e., each point is an estimate of $E[\ln y_{g,m+3} - \ln y_{g,m}]$. For employment retention, each point is a prediction using estimates of equation~(\ref{eq:did_Ret}) for the respective group $g$ and month $m$, i.e., each point is an estimate of $E[\ln E_{g,m+3}]$. \\
\textit{Source:} Earnings Survey, monthly panel January-December 2022, public sector excluded. }
\end{small}
\end{justify}
\end{figure}

Note that one important institutional change was implemented together with the minimum wage increase to \euro{12}, which is that the minijob threshold of originally \euro{450} was harmonized with the minimum wage increase. Hence, with the minimum wage hike in October, the threshold increased to \euro{520}. This expansion of minijobs has been heavily criticized as minijobs are often regarded for offering only precarious job security with few amenities. In addition, it can create an incentive to take on a subsidized minijob instead of looking for a regular job so that the respective individuals end up in the so-called part-time trap \cite{Blomer2022}. However, it addresses the shortcoming highlighted in \citeA{Bossler2023} that the fixed minijob threshold did not allow minijobbers to benefit from a minimum wage increase (while staying in a minijob). Hence, there is no longer an institutional specificity inhibiting the wage effect of the minimum wage. 

Figure \ref{fig:12_minijob_wage} shows hourly and monthly wage effects for regular employees on the left and for minijobbers on the right. For visual clarity of effect sizes, the illustration is restricted to the treatment group and the closest control group because the control groups of minijobbers further up the wage distribution fluctuate a lot.\footnote{Appendix~\ref{app:heterogeneities} contains the full graphical illustration including all control groups.} The hourly wage effect is similar for both groups, which is unsurprising as the workers have been in the same wage group and face the same minimum wage increase. Hence, the similarity of the hourly wage effect does not point to differential non-compliance or varying spillovers (i.e., wage increases beyond the minimum wage). Both groups also experience a monthly wage effect which is slightly smaller in size than the hourly wage increase. 

Figure \ref{fig:12_minijob_employment} displays the working hours and job retention effect for regular employees and minijobbers. While the hours effect is very small (but negative) for regular employees, it is much more emphasized for minijobbers, illustrating a 2 percent decrease. This stronger hours effect of minijobs, although there is no longer the institutional incentive to reduce hours to preserve the minijob, points to a genuine labor demand effect at this vulnerable group of jobs. However, for job retention, there is no such effect heterogeneity. Minijobs have not been destructed in the course of the \euro{12}-minimum wage.

\paragraph{Employment effects at the aggregate.}

So far, we analyzed job retention from the employees' perspective. Examining the individual-level restricts the analysis of an employment effect to job stability since the treatment and control groups can only be defined among workers already employed before the minimum wage was increased. This neglects potential employment effects that operate through the take-up of jobs. Moving the analysis to a more aggregated level, such as regions, allows us to incorporate employment entries. We can identify an overall employment effect if changes in employment vary between regions that are differentially exposed to the minimum wage hike. 

We exploit variation in the regional bite, defined by the fraction of workers paid below the new minimum wage before it came into force. This approach was first proposed by \citeA{Card1992} and used in most previous studies on the German minimum wage introduction, e.g., \citeA{Caliendo2018,Bossler2023}. We use the bite as a treatment variable in a region-level difference-in-differences regression: 
\begin{eqnarray}\label{eq:region}
y_{r,m} = \alpha 
+ \sum_{k =2}^{12} \gamma_{k} M_{k,m}  
+ \sum_{k=2}^{12}  \delta_{k} Bite_{r} * M_{k,m} 
+ \phi Bite_{r} 
+ e_{r,m} 
\\[2ex]
\qquad m = 1, \ldots, 12 \text{ months} \nonumber
\end{eqnarray}
$y_{r,m}$ denotes log employment in region $r$ in month $m$ of the year 2022. The coefficients of interest are the interaction effects of the month dummies $M_{k,m}$ and the bite variable $bite_r$, of which the treatment effects of the minimum wage hike in October 2022 are given by the coefficients $\delta_{10}, \delta_{11}$, and~$\delta_ {12}$. The bite variable is measured in April 2022 using the ES data. See Appendix Figure~\ref{fig:bite} for the distribution of the bite across counties. 

The dependent variable $y_{r,m}$ is adjusted for a bite-specific trend, as in \citeA{Bossler2023} or in \citeA{Dustmann2022}.\footnote{Regression results without adjusting for a bite-specific trend are presented in Appendix Table~\ref{tab:employment_unadj}. However, it is evident that a bite-specific trend is present already before the minimum wage hike.} The bite-specific trend is identified from the following auxiliary regression:
\begin{eqnarray}
y_{r,m} = 
\alpha 
+ \sum_{k =2}^{12} \gamma_{k} M_{k,m}   
+ \tau Bite_{r} *trend_{m} 
+ \sum_{k=10}^{12} \delta_{k} Bite_{r} *M_{k,m} + \phi Bite_{r} + e_{r,m}
\end{eqnarray}
The trend-adjusted dependent variable is then defined as $y_{r,m} - \tau Bite_{r} * trend_{m}$.\footnote{Our analysis window lies in the aftermath of the Covid pandemic (see Appendix~\ref{app:institutional}). Therefore, an employment trend during the year 2022 may arise due to adjustments in the post-Covid period. While regular employment was (due to a generous short-time work policy) barely affected by the pandemic, this was not the case for minijobs, which were indeed rebuilt in 2022. The overall positive (unconditional) development of minijobs is documented in Appendix Table~\ref{tab:app_minijob}.} Standard errors are cluster-robust at the level of counties.

\begin{table}[ht!]
\begin{center}
\caption{Region-level trend-adjusted employment effects}\label{tab:regional_employment_trend}
\resizebox{1.0\linewidth}{!}{
\begin{threeparttable}
\begin{tabular}{lC{1.8cm}C{1.8cm}C{1.8cm}C{1.8cm}C{1.8cm}C{1.8cm}}
\hline \hline
    &\multicolumn{2}{c}{Total employment}&\multicolumn{2}{c}{Minijobs}&\multicolumn{2}{c}{Regular employment} \\
    \cmidrule(lr){2-3}\cmidrule(lr){4-5}\cmidrule(lr){6-7}
    &\multicolumn{1}{c}{(1)}&\multicolumn{1}{c}{(2)}&\multicolumn{1}{c}{(3)}&\multicolumn{1}{c}{(4)}&\multicolumn{1}{c}{(5)}&\multicolumn{1}{c}{(6)}\\
    & \makecell{log} &\makecell{12 months \\log \\difference} & \makecell{log} & \makecell{12 months \\log \\difference} & \makecell{log} & \makecell{12 months \\log \\difference} \\
\hline
$Bite \times January$ & base & base & base & base & base & base \\[0.25ex]
$Bite \times February$ & -0.009***& 0.005  &      -0.025***& -0.009   &      -0.010***&       0.009*  \\
    &     (0.003)         &     (0.006)         &     (0.008)         &     (0.012)         &     (0.003)         &     (0.005)         \\[.25ex]
$Bite \times March$ &      -0.008         &       0.004         &      -0.053***&      -0.040** &      -0.001         &       0.020*  \\
    &     (0.011)         &     (0.011)         &     (0.011)         &     (0.018)         &     (0.012)         &     (0.011)         \\[.25ex]
$Bite \times April$ &       0.004         &       0.023         &      -0.032*  &       0.002          &       0.012         &       0.031** \\
    &     (0.019)         &     (0.017)         &     (0.018)         &     (0.028)         &     (0.019)         &     (0.015)         \\[.25ex]
$Bite \times May$ &      -0.001         &       0.026         &      -0.045*  &       0.014         &       0.008         &       0.030** \\
    &     (0.022)         &     (0.016)         &     (0.024)         &     (0.030)         &     (0.022)         &     (0.015)         \\[.25ex]
$Bite \times June$ &      -0.004         &       0.018         &      -0.033         &       0.009         &       0.011         &       0.023*  \\
    &     (0.026)         &     (0.013)         &     (0.029)         &     (0.032)         &     (0.023)         &     (0.012)         \\[.25ex]
$Bite \times July$ &       0.002         &       0.009         &      -0.022         &      -0.028         &       0.003         &       0.013         \\
    &     (0.030)         &     (0.014)         &     (0.043)         &     (0.039)         &     (0.025)         &     (0.013)         \\[.25ex]
$Bite \times August$ &       0.013         &       0.004         &      -0.008         &      -0.024         &       0.001         &       0.010         \\
    &     (0.034)         &     (0.020)         &     (0.050)         &     (0.045)         &     (0.030)         &     (0.020)         \\[.25ex]
$Bite \times September$ &      -0.020         &      -0.000         &      -0.028         &       0.003         &      -0.011         &       0.005         \\
                    &     (0.032)         &     (0.022)         &     (0.048)         &     (0.050)         &     (0.030)         &     (0.022)         \\[.25ex]
$Bite \times October$ &      -0.033         &      -0.007         &      -0.083*  &       0.008         &      -0.030         &       0.001         \\
                    &     (0.030)         &     (0.023)         &     (0.045)         &     (0.050)         &     (0.029)         &     (0.023)         \\[.25ex]
$Bite \times November$ &      -0.052** &      -0.007         &      -0.128***&       0.004         &      -0.043*  &       0.006         \\
                    &     (0.025)         &     (0.023)         &     (0.043)         &     (0.052)         &     (0.025)         &     (0.023)         \\[.25ex]
$Bite \times December$ &      -0.074***&      -0.012         &      -0.170***&       0.004         &      -0.064***&       0.003         \\
                    &     (0.023)         &     (0.023)         &     (0.043)         &     (0.054)         &     (0.023)         &     (0.023)         \\
\hline
Cluster        &        400         &        400         &        400         &        400         &        400         &        400         \\
Observations        &        4800         &        4800         &        4800         &        4800         &        4800         &        4800         \\
\hline \hline
\end{tabular}
\begin{small} \textit{Notes:} Treatment interactions from region-level difference-in-difference regressions as specified in equation (\ref{eq:region}). Outcome variables are the trend-adjusted logarithm and, respectively, the trend-adjusted 12-month difference of the logarithm of total employment, minijobs, and regular social security employees. Standard errors in parentheses are clustered at the county level. Asterisks indicate significance levels *~$p<0.05$, **~$p<0.01$, ***~$p<0.001$. 
\\ \textit{Source:} County-level administrative employment data of the Federal Employment Agency, and Earnings Survey of April 2022 for the regionally aggregated bite. 
\end{small}
\end{threeparttable}
}
\end{center}
\end{table}

In a second employment specification, we not only correct for a bite-specific trend but also adjust for seasonality in employment by taking the 12-month log-difference, i.e., we compare log employment to the respective value in the year 2021: $y_{r,m} - y_{r,m-12}$. This approach is based on the understanding that employment follows a particular seasonality and it is feasible since the administrative county-level data allow us to go further back in time than the year 2022.

Table~\ref{tab:regional_employment_trend} reports the interaction effects of the employment regressions.\footnote{A full table including all regression coefficients is presented in Appendix Table~\ref{tab:employment_full_table}.} The results on total employment are displayed in the first two columns, according to which any employment changes until September are irrespective of the regional bite, i.e., the respective coefficients are close to zero until September. In the final months (after the minimum wage hike), the estimates turn negative in column (1), pointing to a small negative region-level employment effect of the minimum wage. However, when accounting for seasonality in column (2), the respective treatment effects turn insignificant and shrink towards zero. We observe similar effect patterns when analyzing minijobs in columns (3) and (4), as well as regular employment in columns (5) and (6). While estimations for both employment groups point to a negative treatment effect when using log employment as the dependent variable (columns (3) and (5)), they are virtually zero in the regressions that account for seasonality (columns (4) and (6)). To summarize, incorporating employment entries into the analysis has not changed the findings from the individual-level job retention regressions, and we conclude that the 2022 minimum wage hike has neither reduced job stability nor the aggregate employment level.

\paragraph{Effects using the wage gap.}

Instead of classifying individuals into distinct wage groups defining their treatment status, we now use an alternative treatment definition, the so-called \textit{bite gap}, defining each individual's wage gap to the forthcoming minimum wage level of \euro{12}. I.e., the \textit{bite gap} measures the percentage wage increase of employee $i$ that is required by the October 2022 minimum wage hike: 
\begin{align}
\textit{Bite gap}_{i,m} = \begin{cases} (\textit{\euro}{12}-w_{i,m})/w_{i,m} &\text{if } w_{i,m}<\textit{\euro}12 \\ 0 &\text{if } w_{i,m}\geq\textit{\euro}12 
\end{cases}
\end{align}

Table \ref{tab:gap_variable} reports the \textit{bite gap} for each month of 2022. The minimum wage increase to \euro{12} required a wage increase in the range of two to three percent (as of January to September) across all low-wage workers in the wage range between 10.45 and 14.99. Even more importantly, the minimum wage required an effective wage increase of six to eight percent for the treated employees.\footnote{We define the \textit{bite gap} only among the groups of treated employees but not among employees who are already affected by the previous minimum wage hike to \euro{10.45} which eases comparability with the results presented so far. We also exclude employees paid non-compliant before the minimum wage increases in 2022. Since many of the non-compliant workers do not receive a pay rise (see Table~\ref{tab:app_baselinereg}), they would be assigned a high bite without a corresponding wage effect. Consequently, the respective workers are excluded from the bite gap analysis.} Note that the bite gap fell in October when the minimum wage increased, indicating its effectiveness in lifting wages. 
%The first minimum wage increase in July also led to a narrowing of the wage gap but to a smaller degree. Among treated employees, the bite gap remains fairly constant up to September, indicating that the treatment group was relatively stable before the \euro{12} minimum wage hike. 

\begin{table}[ht!]
\begin{center}
\caption{Average bite gap} \label{tab:gap_variable}
\resizebox{1.00\linewidth}{!}{
\begin{threeparttable}
\begin{tabular}{l ccc ccc ccc ccc}
\hline \hline
 & Jan.  &   Feb.&  March&   April&   May   &  June &   July  &  Aug.&   Sept.&   Oct.&   Nov.&   Dec. \\[1ex]
\hline
\makecell{Of employees with wage ...} \\[1ex]  
\quad $10.45 - 14.99$ 
& 0.022 &	0.022 &	0.022 &	0.021 &	0.021 &	0.021 &	0.031 &	 0.030 &	0.030 &	0.005 &	0.004 &	0.003
\\[2ex]
\quad  $10.45 - 11.99$  &  
0.062 &	0.063 &	0.065 &	0.062 &	0.063 &	0.063 &	0.078 &	0.077 &	0.077 &	0.039 &	0.035 &	 0.030 \\
\hline \hline
\end{tabular}
\begin{small} 
\textit{Notes:} Figures are based on weighted data using the simple inverse probability weighting factor. \\
\textit{Source:} Earnings Survey, monthly panel January-December 2022, public sector and employees with an hourly wage of less than \euro{10.45} and more than \euro{14.99} are excluded. \raggedright
\end{small}
\end{threeparttable}
}
\end{center}
\end{table}

The regression specification identifies treatment effects from an interaction of the bite gap with months dummies: 
\begin{eqnarray}\label{eq:bite_gap}
\ln y_{i,m+3} - \ln y_{i,m} = 
\alpha 
+ \sum_{k =2}^{9} \delta_{k} M_{k,m} * \textit{Bite Gap}_{i,m} 
+ \sum_{k=2}^{9} \gamma_{k} M_{k,m}  
+ \phi \textit{Bite Gap}_{i,m}  
+ v_{i,m} \\[2ex]
\qquad \qquad m = 1, \ldots, 9 \text{ months} \nonumber
\end{eqnarray}

The use of the bite gap has the advantage of capturing treatment heterogeneity within the treatment group, as we expect a larger wage effect for workers with a large gap. In fact, the increase in hourly wages should --by and large-- correspond with the \textit{bite gap}, that is, we would expect a treatment effect of one if the wage gap is fully closed by the minimum wage.\footnote{In fact, the bite gap may not only be closed by the minimum wage. It may also be closed by the common monthly time effects $\alpha + \gamma_{k}$, which capture wage growth that occurs irrespective of the bite gap, as well as by a time-constant bite gap effect $\phi$ that occurs irrespective of the minimum wage hikes and captures mean-reversion effects.}

\begin{table}[ht!]
\begin{center}
\def\sym#1{\ifmmode^{#1}\else\(^{#1}\)\fi}
\caption{Treatment effects of the wage gap} \label{tab:gap_regression}
\resizebox{1.0\linewidth}{!}{
\begin{threeparttable}
\begin{tabular}{l *{4}{d{6}l}} 
\hline \hline
&\multicolumn{2}{c}{(1)} 
&\multicolumn{2}{c}{(2)} 
&\multicolumn{2}{c}{(3)} 
&\multicolumn{2}{c}{(4)} \\
&\multicolumn{2}{c}{\text{$\Delta \ln$ hourly wage}} 
&\multicolumn{2}{c}{$\Delta \ln$ monthly wage} 
&\multicolumn{2}{c}{$\Delta \ln$ monthly} 
&\multicolumn{2}{c}{Employed in same}\\
& & &  & & \multicolumn{2}{c}{contractual hours}
&\multicolumn{2}{c}{plant in $t+3$ } \\
& & & & & & & \multicolumn{2}{c}{(Dummy)} \\[0.5ex]
\hline
\textit{Bite Gap}    &  0.1334\sym{**}   &(0.0202)& 0.2015\sym{***}   &(0.0435)&  0.0682           &(0.0424)&-0.5624\sym{***}   &(0.0477)  \\[0.5ex]
\ $\times January$   & \multicolumn{2}{c}{base} & \multicolumn{2}{c}{base}&\multicolumn{2}{c}{base} & \multicolumn{2}{c}{base}\\
\ $\times February$  &   0.0470          &(0.0266)& 0.0399            &(0.0520)& -0.0071           &(0.0519)& 0.0844            &(0.0549)  \\
\ $\times March$     &  -0.0036          &(0.0276)&-0.1333\sym{*}     &(0.0534)& -0.1297\sym{*}    &(0.0515)& 0.0513            &(0.0610)  \\
\ $\times April$     &   0.0891\sym{**}  &(0.0278)& 0.0416            &(0.0610)& -0.0475           &(0.0609)& 0.1338\sym{*}     &(0.0663)  \\
\ $\times May$       &   0.1125\sym{***} &(0.0277)&-0.0323            &(0.0573)& -0.1448\sym{*}    &(0.0573)& 0.0799            &(0.0663)  \\
\ $\times June$      &   0.0728\sym{**}  &(0.0273)&-0.0521            &(0.0637)& -0.1250\sym{*}    &(0.0614)& 0.0887            &(0.0690)  \\
\ $\times July$      &   0.6577\sym{***} &(0.0222)& 0.4040\sym{***}   &(0.0558)& -0.2537\sym{***}  &(0.0544)& 0.1001            &(0.0595)  \\
\ $\times August$    &   0.5974\sym{***} &(0.0253)& 0.3758\sym{***}   &(0.0603)& -0.2216\sym{***}  &(0.0579)& 0.1395\sym{*}     &(0.0597)  \\
\ $\times September$ &   0.6989\sym{***} &(0.0307)& 0.5206\sym{***}   &(0.0651)& -0.1783\sym{**}   &(0.0589)& 0.0342            &(0.0610)  \\[0.5ex]
$January$            & \multicolumn{2}{c}{base} & \multicolumn{2}{c}{base}&\multicolumn{2}{c}{base} & \multicolumn{2}{c}{base}            \\
$February$           &  -0.0101\sym{***} &(0.0008)&  0.0188\sym{***}  &(0.0017)&  0.0288\sym{***}  &(0.0017)& 0.0075\sym{***}   &(0.0020)  \\
$March$              &   0.0023\sym{*}   &(0.0009)&-0.0161\sym{***}   &(0.0019)& -0.0185\sym{***}  &(0.0020)& 0.0091\sym{***}   &(0.0023)  \\
$April$              &  -0.0042\sym{***} &(0.0009)&-0.0113\sym{***}   &(0.0020)& -0.0071\sym{***}  &(0.0021)& 0.0200\sym{***}   &(0.0025)  \\
$May$                &  -0.0074\sym{***} &(0.0009)&-0.0104\sym{***}   &(0.0021)& -0.0030           &(0.0021)& 0.0149\sym{***}   &(0.0025)  \\
$June$               &   0.0030\sym{**}  &(0.0010)&-0.0055\sym{*}     &(0.0022)& -0.0085\sym{***}  &(0.0021)&-0.0065\sym{*}     &(0.0026)  \\
$July$               &   0.0240\sym{***} &(0.0009)& 0.0121\sym{***}   &(0.0020)& -0.0119\sym{***}  &(0.0020)& 0.0081\sym{***}   &(0.0024)  \\
$August$             &   0.0562\sym{***} &(0.0012)& 0.0357\sym{***}   &(0.0021)& -0.0206\sym{***}  &(0.0020)& 0.0015            &(0.0025)  \\
$September$          &   0.0494\sym{***} &(0.0014)& 0.0266\sym{***}   &(0.0027)& -0.0228\sym{***}  &(0.0024)&-0.0034            &(0.0026)  \\[0.5ex]
Constant             &   0.0272\sym{***} &(0.0007)& 0.0381\sym{***}   &(0.0014)&  0.0109\sym{***}  &(0.0015)& 0.8312\sym{***}   &(0.0018)  \\[0.5ex]      
\hline
Employees    & \multicolumn{2}{c}{2,053,133} & \multicolumn{2}{c}{2,053,133} & \multicolumn{2}{c}{2,053,133}  & \multicolumn{2}{c}{2,569,311}     \\ 
Observations & \multicolumn{2}{c}{9,121,144} & \multicolumn{2}{c}{9,121,144} & \multicolumn{2}{c}{9,121,144}
& \multicolumn{2}{c}{10,972,592} \\
\hline \hline
\end{tabular}
\begin{small} \textit{Notes:} Individual-level difference-in-difference regressions as specified in equation (\ref{eq:bite_gap}). Outcome variables are the natural logarithm of (1) hourly wages, (2) monthly wages, (3) working hours, and (4) a binary variable equal to $1$ if the employee works in the same plant in $t+3$ and $t$. 
All regressions are weighted. Standard errors in parentheses are clustered at the individual level. Asterisks indicate significance levels. *~$p<0.05$, **~$p<0.01$, ***~$p<0.001$.\\
\textit{Source:} Earnings Survey, monthly panel January-December 2022, only employees with gross hourly wages between $10.45$ and $14.99$ in period $t$, public sector excluded. 
\end{small}
\end{threeparttable}
}
\end{center}
\end{table}

The results of column (1) in Table~\ref{tab:gap_regression} show a significant hourly wage response to the minimum wage adjustment in October. More precisely, in the months that capture the 3-month-wage growth across the minimum wage increase (\textit{July, August, and September}), the effect lies in the range between 0.60 and 0.70. Together with the baseline monthly wage development and with the time-constant bite effect, the effect size indicates that the initial gap almost closed (e.g., for wage growth up to December, which is indicated by the September effects: $0.699 + 0.049 + 0.133 + 0.027 = 0.908$), leaving some scope for non-compliance. That is, the result suggests that the remaining wage gap, which indirectly measures non-compliance, is below 10 percent among the affected employees. If, however, wages of some employees are raised beyond the minimum wage (resulting in a larger wage effect due to spillovers) other individuals' wages may fall short of the minimum wage, which further increases non-compliance. The estimates also show that the wage gap already partly closed when the minimum wage was raised to \euro{10.45}, as indicated by the wage growth effects of the months \textit{April, May, and June}, indicating that about 10\% of the wage gap was closed by this initial policy change. 

The effect in column (2) of Table~\ref{tab:gap_regression} shows a large increase also in monthly wages. Still, the effect is significantly smaller than the hourly wage effect, again supporting a modest negative effect on working hours. The effect on working hours is presented in column (3). The treatment effects, as indicated by the interaction effects of July, August, and September, are significantly negative and correspond in size with the baseline effects of Figure~\ref{fig:12_base_employment}. However, the results show a negative path already before the minimum wage was raised, indicating that the initial minimum wage increase to \euro{10.45} in July 2022 plays a role. We will elaborate on the effects of this preceding hike in the next subsection. Finally, column (4) of Table~\ref{tab:gap_regression} does not point to a treatment effect on employment retention in the months after the minimum wage increase. 
%If anything, the treated workers over the months of 2022 more likely preserved their jobs, which may be caused by a recovery in the aftermath of the Covid-19 pandemic rather than a genuine minimum wage effect in October 2022. 

\paragraph{Effects of the \euro{10.45} minimum wage.}

\begin{figure}[ht!]
\centering
\caption{Hourly and monthly wage effect of \euro{10.45} minimum wage}\label{fig:1045_wage}
\includegraphics[width=0.49\textwidth]{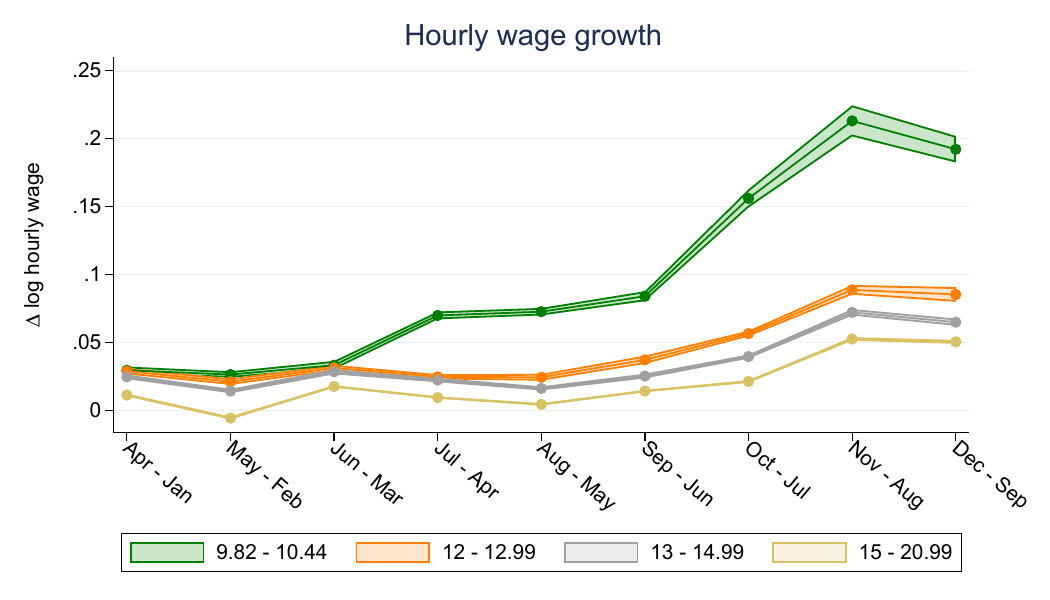}
\includegraphics[width=0.49\textwidth]{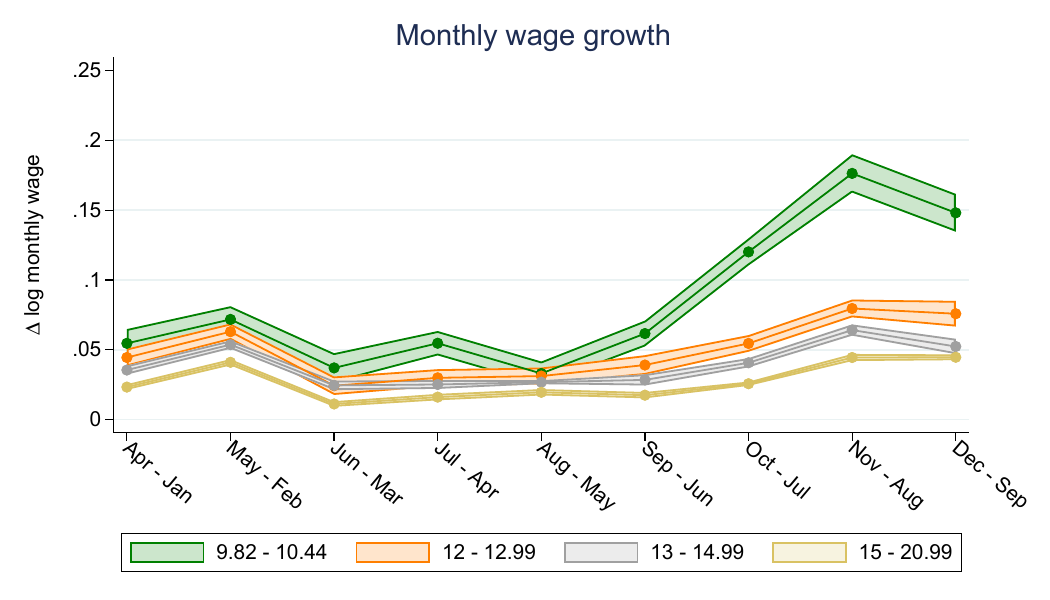}
\begin{justify}
\begin{small}{\textit{Notes:} The left part presents the development of hourly wage growth and the right part presents the development of monthly wage growth. Each point is a prediction using the estimates of equation~(\ref{eq:didid}) for the respective group $g$ and month $m$, i.e., each point is an estimate of $E[\ln y_{g,m+3} - \ln y_{g,m}]$. Shaded areas show the 95\%-confidence intervals with standard errors clustered at the individual level.\\
\textit{Source:} Earnings Survey, monthly panel January-December 2022, public sector excluded. }
\end{small}
\end{justify}
\end{figure}

To analyze the group of workers that have already been affected by the \euro{10.45} minimum wage increase, we illustrate the results for the respective group ($ {\mbox {\euro}} 9.82 \leq w_{i,m} \leq {\mbox {\euro}}10.44$), as specified in equations~(\ref{eq:didid}) and~(\ref{eq:did_Ret}). Figure~\ref{fig:1045_wage} presents the treatment effects for hourly and monthly wage growth for the treated (green line) along with the same control groups as in Figure~\ref{fig:12_base_wage}. The first three data points reflect pre-treatment periods. Since the hourly and monthly wage growth evolved very similarly between the treatment and the control groups, we can be confident about the identifying parallel trends-assumption. The following three data points (\textit{July-April}, \textit{August-May}, \textit{September-June}) cover the minimum wage increase of July 1st, thereby capturing the treatment effect of the \euro{10.45}-minimum wage hike. All three data points clearly indicate that there is a positive wage effect of about 5 percent. Interestingly, the monthly wage effect is less clear. If anything, it is small, and it definitely falls short of the hourly wage response, suggesting a negative working hours adjustment. The last three data points of the green line reflect a lagged effect of a relatively small group of workers that was already affected by the minimum wage increase in July but remained in the respective group (because of non-compliance). However, the visual inspection makes it obvious that the respective workers received, on average, a pronounced lagged wage rise when the \euro{12}-minimum wage came into force. This latter result suggests that the \euro{12}-minimum wage was much more prominent (and publicly debated) and therefore more effective than the smaller increase in July.

\begin{figure}[ht!]
\centering
\caption{Working hours and employment retention effect of \euro{10.45} minimum wage} \label{fig:1045_employment}
\includegraphics[width=0.49\textwidth]{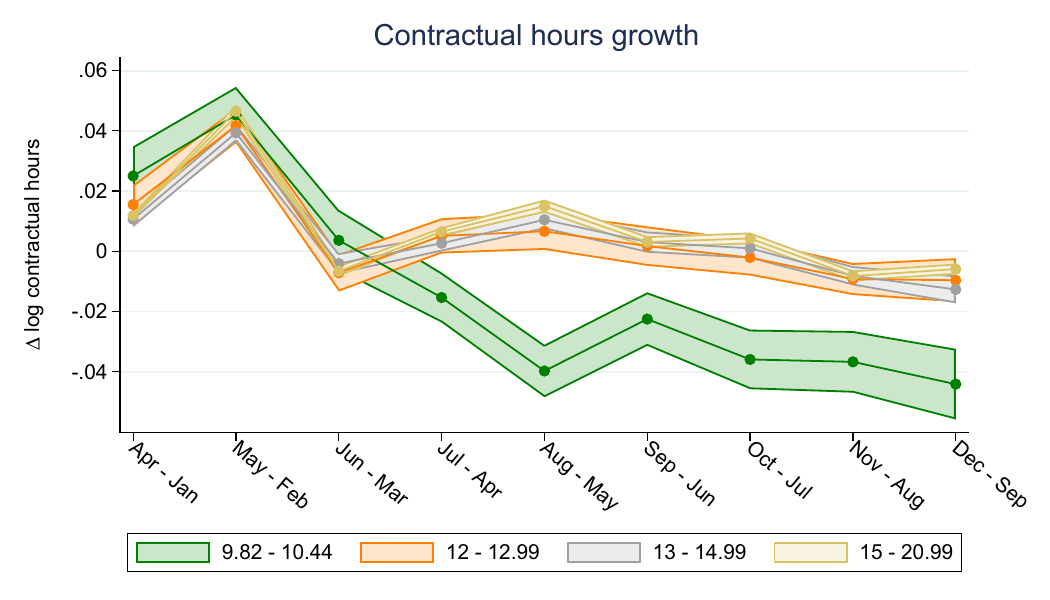}
\includegraphics[width=0.49\textwidth]{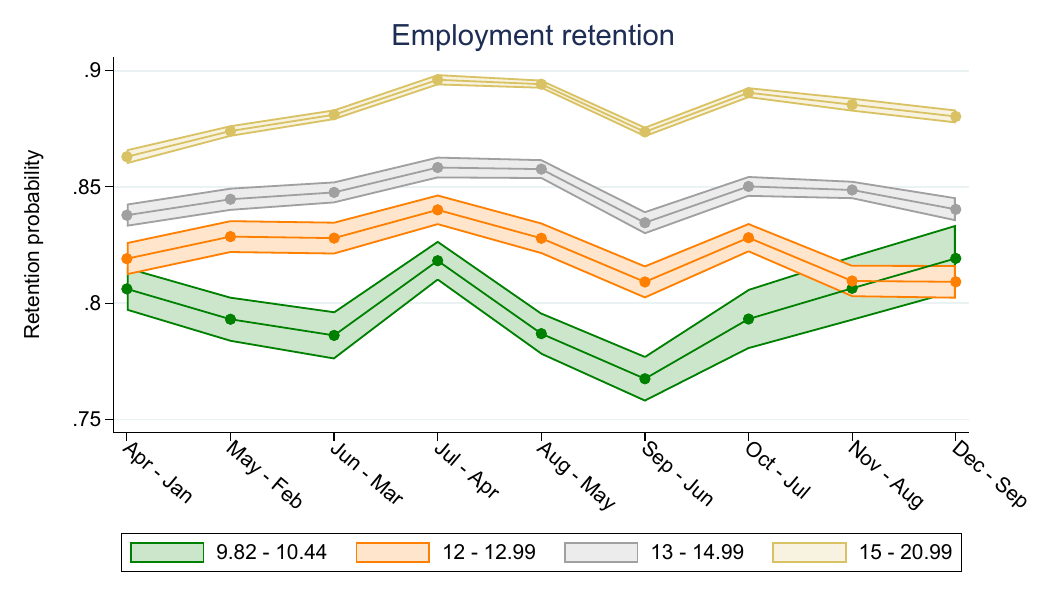} 
\begin{justify}
\begin{small}
{\textit{Notes:} The left part presents the development of monthly working hours growth and the right part presents the development of employment retention. For monthly working hours, each point is a prediction using the estimates of equation~(\ref{eq:didid}) for the respective group $g$ and month $m$, i.e., each point is an estimate of $E[\ln y_{g,m+3} - \ln y_{g,m}]$. For employment retention, each point is a prediction using estimates of equation~(\ref{eq:did_Ret}) for the respective group $g$ and month $m$, i.e., each point is an estimate of $E[\ln E_{g,m+3}]$. Shaded areas show the 95\%-confidence intervals with standard errors clustered at the individual level.\\ \textit{Source:} Earnings Survey, monthly panel January-December 2022, public sector excluded. }
\end{small}
\end{justify}
\end{figure}

Figure \ref{fig:1045_employment} shows the negative hours response when the \euro{10.45}-minimum wage hike took place, in size corresponding with the difference between the hourly and monthly wage effect. However, there is no increase or decrease in employment retention before October. The last three data points, though, demonstrate a strong positive effect on employment retention, indicating that employees who were initially paid non-compliant but experienced a significantly positive lagged increase in hourly and monthly wages also observed a lagged increase in their likelihood of job retention. This emphasizes the positive lagged effect for a group of workers who were initially paid non-compliant. Note, however, that the initial hourly wage effect was almost completely crowded out by the negative hours effect.

\section{Conclusion}
\label{sec:conclusion}

We are the first to analyze the labor market effects of the minimum wage increase to \euro{12} in October 2022, which made the German minimum wage the highest PPP-adjusted minimum wage across the European Union. We exploit novel data on a large number of employees, which is collected monthly by the German Statistical Office, allowing us to classify individuals into narrowly defined treatment and control groups. We estimate individual-level difference-in-differences-in-differences specifications that allow us to control for mean-reversion in wages. 

The results demonstrate unambiguous positive effects on hourly wages. The effect size suggests that the affected employees' wage gap to the new minimum wage is almost fully closed, leaving only little room for non-compliance. However, in the short run, we observe a small downward adjustment in working hours among treated employees. This hours response partially offsets the effect in terms of monthly wages, which is still positive when looking at the minimum wage increase to \euro{12}. 

Regarding employment effects in heads, we neither identify an effect on the workers' job retention nor on the region-level employment of total jobs, regular jobs, or minijobs. Hence, all the employment responses come from the intensive margin of working hours. The final minimum wage increase to \euro{12} implies a labor demand elasticity of $-0.2$, which is relatively modest in size. 

%When also accounting for the hours effect of the penultimate minimum wage increase, the overall labor demand elasticity amounts to $-0.33$. 

%%%%%%%%%%%%%%%%%%%%%%%%%%%%%%%%%%%%%%%%%%%%%%%%%
\clearpage
\begin{onehalfspacing}
\bibliographystyle{chicago}
\bibliography{manuscript.bib}
\end{onehalfspacing}
%%%%%%%%%%%%%%%%%%%%%%%%%%%%%%%%%%%%%%%%%%%%%%%%%

\clearpage

\vspace{1cm}

\begin{center}

\Large
Online Appendix for \\

\vspace{1cm}

A 22 percent increase in the German minimum wage: \\ nothing crazy!

\normalsize
\vspace{1cm}

by \\
Mario Bossler, Lars Chittka, and Thorsten Schank

\vspace{1.0cm}
\Large 
\textbf{Content}
\normalsize

\singlespacing
\startcontents[sections] 
\printcontents[sections]{l}{1}{\setcounter{tocdepth}{2}} 

\end{center}

\clearpage

\begin{appendices}

\setcounter{figure}{0}
\setcounter{table}{0}
\renewcommand*{\thefigure}{\thesection\arabic{figure}}
\renewcommand*{\thetable}{\thesection\arabic{table}}
\section{Institutional background}
\label{app:institutional}

\begin{figure}[ht!]
\centering
\caption{Minimum Wages in the European Union. }\label{fig:Europe}
\includegraphics[width=0.7\textwidth]{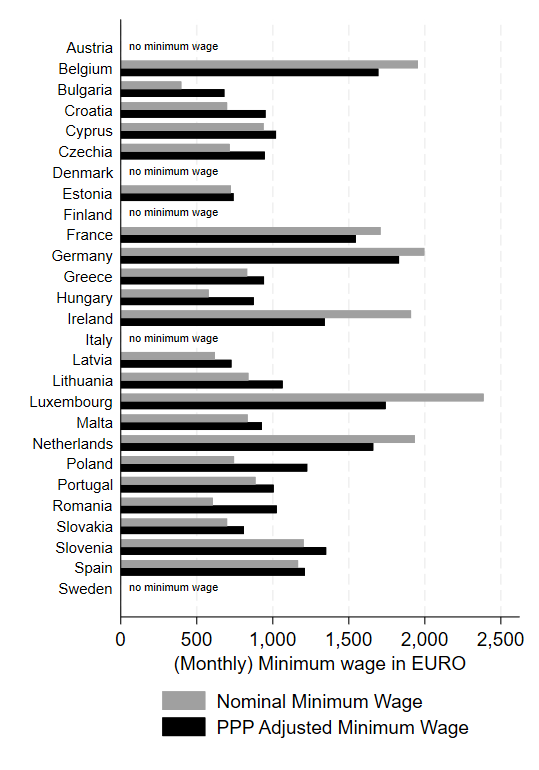}
\begin{justify}
\begin{small}{%\textit{Notes:} 
\textit{Source:} Eurostat, first half of 2023. }
\end{small}
\end{justify}
\end{figure}

\begin{figure}[ht!]
\centering
\caption{Google search requests regarding the corona pandemic. }\label{fig:Google}
\includegraphics[width=0.98\textwidth]{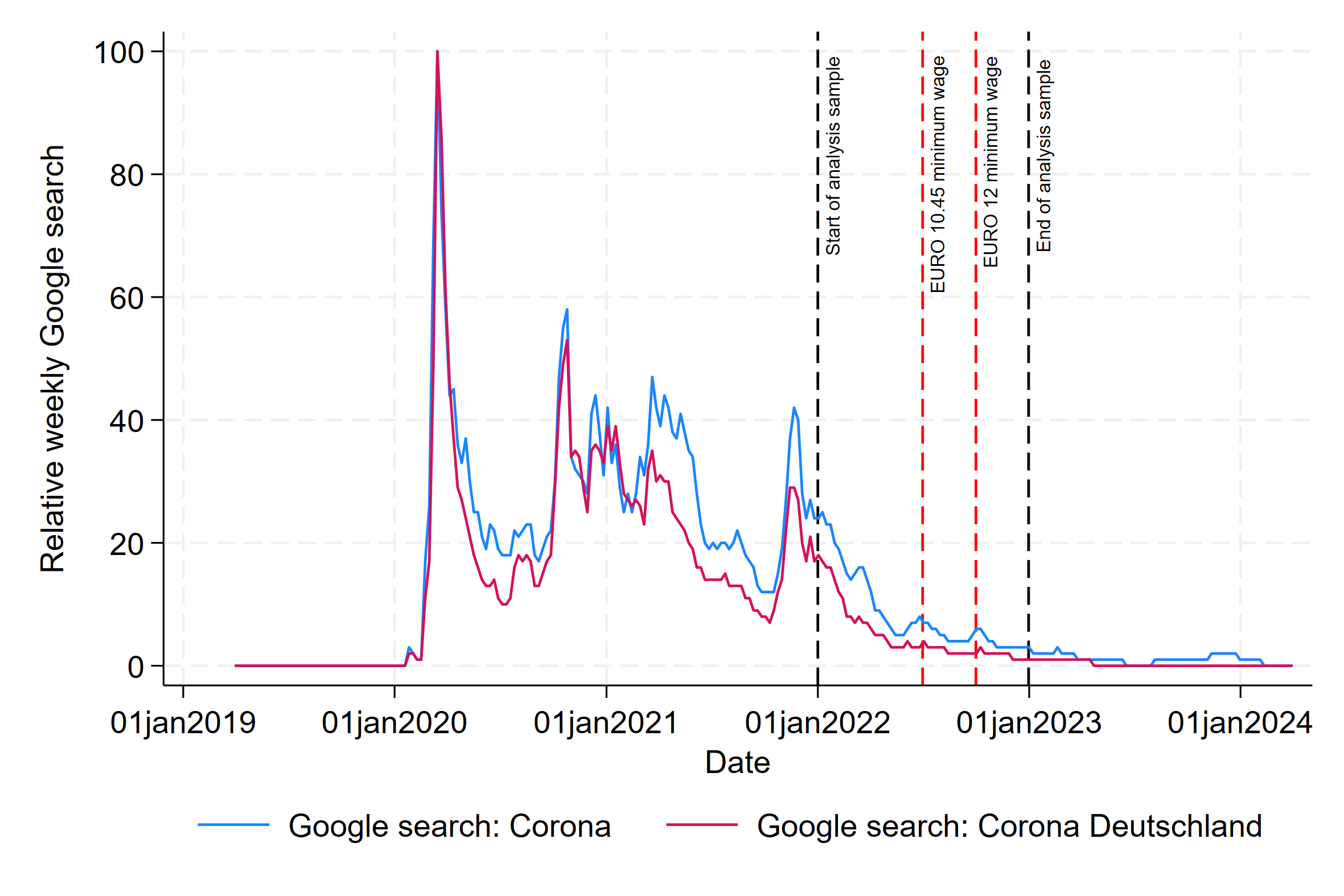}
\begin{justify}
\begin{small} {%\textit{Notes:} 
\textit{Source:} Google Trends, data retrieved on April 2nd 2024. }
\end{small}
\end{justify}
\end{figure}

\clearpage
\setcounter{figure}{0}
\setcounter{table}{0}
\renewcommand*{\thefigure}{\thesection\arabic{figure}}
\renewcommand*{\thetable}{\thesection\arabic{table}}
\section{Baseline regression results}
\label{app:reg_base}

\begin{table}[ht!]\centering
\caption{Regression coefficients of baseline regressions}
\label{tab:app_baselinereg}
\begin{footnotesize}
\begin{threeparttable}
\renewcommand{\arraystretch}{1.1}
\begin{tabular}{l D{.}{.}{1.5}l D{.}{.}{1.5}l D{.}{.}{1.5}l D{.}{.}{1.5}l}
\hline\hline
&\multicolumn{2}{c}{(1)} 
&\multicolumn{2}{c}{(2)} 
&\multicolumn{2}{c}{(3)} 
&\multicolumn{2}{c}{(4)} \\
&\multicolumn{2}{c}{\text{$\Delta \ln$ hourly wage}} 
&\multicolumn{2}{c}{$\Delta \ln$ monthly wage} 
&\multicolumn{2}{c}{\text{$\Delta \ln$ monthly}} 
&\multicolumn{2}{c}{Employed in same}\\
& & & & & \multicolumn{2}{c}{contractual hours} 
&\multicolumn{2}{c}{plant in $t+3$ (Dummy) } \\
\hline
Constant                             &  0.0289^{***} & (0.0012)&  0.0444^{***} & (0.0032)& 0.0155^{***}  &(0.0033) & 0.8191^{***}  & (0.0035) \\
May -- February                      &  -0.0077^{***}& (0.0013)&  0.0185^{***} & (0.0036)& 0.0262^{***}  &(0.0036) & 0.0095^{*}    & (0.0037) \\
June -- March                        &  0.0024       & (0.0016)&  -0.0202^{***}& (0.0042)& -0.0227^{***} &(0.0041) & 0.0089^{*}    & (0.0043) \\
July -- April                        &  -0.0042^{**} & (0.0015)&  -0.0146^{**} & (0.0046)& -0.0104^{*}   &(0.0047) & 0.0210^{***}  & (0.0047) \\
August -- May                        &  -0.0045^{**} & (0.0017)&  -0.0133^{**} & (0.0046)& -0.0089       &(0.0046) & 0.0088        & (0.0048) \\
September -- June                    &  0.0084^{***} & (0.0019)&  -0.0054      & (0.0047)& -0.0138^{**}  &(0.0047) & -0.0100^{*}   & (0.0049) \\
October -- July                      &  0.0277^{***} & (0.0015)&  0.0101^{*}   & (0.0042)& -0.0176^{***} &(0.0043) & 0.0090        & (0.0046) \\
November -- August                   &  0.0599^{***} & (0.0021)&  0.0351^{***} & (0.0045)& -0.0247^{***} &(0.0043) & -0.0096^{*}   & (0.0048) \\
December -- September                &  0.0565^{***} & (0.0030)&  0.0314^{***} & (0.0056)& -0.0251^{***} &(0.0050) & -0.0099^{*}   & (0.0049) \\[1.5ex]

\multicolumn{2}{l}{\textbf{Gross hourly wage $<$ \euro{7.50}}} \\
\quad $\times$ April -- January      &  0.5956^{***} & (0.0091)&  0.3226^{***} & (0.0125)& -0.2730^{***} &(0.0134) & -0.1328^{***} & (0.0087) \\
\quad $\times$ May -- February       &  0.5821^{***} & (0.0107)&  0.3057^{***} & (0.0181)& -0.2766^{***} &(0.0176) & -0.1337^{***} & (0.0088) \\
\quad $\times$ June -- March         &  0.5620^{***} & (0.0106)&  0.2799^{***} & (0.0141)& -0.2821^{***} &(0.0153) & -0.1386^{***} & (0.0088) \\
\quad $\times$ July -- April         &  0.3407^{***} & (0.0109)&  0.2327^{***} & (0.0233)& -0.1079^{***} &(0.0223) & -0.1353^{***} & (0.0126) \\
\quad $\times$ August -- May         &  0.3538^{***} & (0.0145)&  0.2380^{***} & (0.0223)& -0.1158^{***} &(0.0186) & -0.1529^{***} & (0.0125) \\
\quad $\times$ September -- June     &  0.3641^{***} & (0.0135)&  0.2405^{***} & (0.0161)& -0.1236^{***} &(0.0137) & -0.1201^{***} & (0.0118) \\
\quad $\times$ October -- July       &  0.3411^{***} & (0.0129)&  0.2338^{***} & (0.0171)& -0.1073^{***} &(0.0159) & -0.1399^{***} & (0.0123) \\
\quad $\times$ November -- August    &  0.3529^{***} & (0.0124)&  0.2384^{***} & (0.0166)& -0.1144^{***} &(0.0154) & -0.1223^{***} & (0.0126) \\
\quad $\times$ December -- September &  0.3539^{***} & (0.0181)&  0.2247^{***} & (0.0212)& -0.1292^{***} &(0.0215) & -0.1354^{***} & (0.0146) \\[1.5ex]

\multicolumn{2}{l}{\textbf{Gross hourly wage \euro{7.50 --9.81}}} \\
\quad $\times$ April -- January      &  0.1337^{***} & (0.0057)&  0.0893^{***} & (0.0094)& -0.0444^{***} &(0.0086) & -0.1867^{***} & (0.0081) \\
\quad $\times$ May -- February       &  0.1129^{***} & (0.0042)&  0.0698^{***} & (0.0069)& -0.0432^{***} &(0.0074) & -0.1083^{***} & (0.0084) \\
\quad $\times$ June -- March         &  0.0966^{***} & (0.0036)&  0.0567^{***} & (0.0057)& -0.0399^{***} &(0.0061) & -0.0384^{***} & (0.0072) \\
\quad $\times$ July -- April         &  0.1346^{***} & (0.0045)&  0.0858^{***} & (0.0066)& -0.0488^{***} &(0.0066) & -0.0856^{***} & (0.0078) \\
\quad $\times$ August -- May         &  0.1305^{***} & (0.0046)&  0.0815^{***} & (0.0085)& -0.0490^{***} &(0.0080) & -0.0470^{***} & (0.0073) \\
\quad $\times$  September -- June    &  0.1412^{***} & (0.0053)&  0.0996^{***} & (0.0077)& -0.0415^{***} &(0.0072) & -0.0593^{***} & (0.0082) \\
\quad $\times$ October -- July       &  0.1863^{***} & (0.0065)&  0.1232^{***} & (0.0107)& -0.0631^{***} &(0.0108) & -0.0836^{***} & (0.0092) \\
\quad $\times$ November -- August    &  0.1891^{***} & (0.0072)&  0.1486^{***} & (0.0092)& -0.0405^{***} &(0.0082) & -0.0486^{***} & (0.0089) \\
\quad $\times$ December -- September &  0.1890^{***} & (0.0097)&  0.1216^{***} & (0.0127)& -0.0673^{***} &(0.0087) & -0.0963^{***} & (0.0106)  \\[1.5ex]

\multicolumn{2}{l}{\textbf{Gross hourly wage: \euro{9.82 --10.45}}} \\
\quad $\times$ April -- January      &  0.0007       & (0.0018)&  0.0102       & (0.0060)& 0.0095        &(0.0060) & -0.0130^{*}   & (0.0058) \\
\quad $\times$ May -- February       &  0.0053^{**}  & (0.0017)&  0.0088       & (0.0055)& 0.0035        &(0.0055) & -0.0355^{***} & (0.0060) \\
\quad $\times$ June -- March         &  0.0020       & (0.0019)&  0.0128^{*}   & (0.0062)& 0.0108        &(0.0059) & -0.0418^{***} & (0.0062) \\
\quad $\times$ July -- April         &  0.0452^{***} & (0.0017)&  0.0248^{***} & (0.0052)& -0.0205^{***} &(0.0051) & -0.0219^{***} & (0.0053) \\
\quad $\times$ August -- May         &  0.0482^{***} & (0.0018)&  0.0019       & (0.0055)& -0.0464^{***} &(0.0052) & -0.0410^{***} & (0.0056) \\
\quad $\times$ September -- June     &  0.0468^{***} & (0.0024)&  0.0226^{***} & (0.0057)& -0.0242^{***} &(0.0055) & -0.0416^{***} & (0.0060) \\
\quad $\times$ October -- July       &  0.0994^{***} & (0.0035)&  0.0656^{***} & (0.0057)& -0.0338^{***} &(0.0057) & -0.0350^{***} & (0.0071) \\
\quad $\times$ November -- August    &  0.1243^{***} & (0.0060)&  0.0968^{***} & (0.0075)& -0.0275^{***} &(0.0057) & -0.0031       & (0.0077) \\
\quad $\times$ December -- September &  0.1068^{***} & (0.0056)&  0.0723^{***} & (0.0082)& -0.0345^{***} &(0.0070) & 0.0101        & (0.0080)  \\[1.5ex]             
\hline
\end{tabular}
\raggedleft{Continued overleaf}
\end{threeparttable}
\end{footnotesize}
\end{table}

\setcounter{table}{0}
\begin{table}[ht!]\centering
\caption{Continued}
\begin{footnotesize}
\begin{threeparttable}
\renewcommand{\arraystretch}{1.1}
\begin{tabular}{lD{.}{.}{1.5}l D{.}{.}{1.5}l D{.}{.}{1.5}l D{.}{.}{1.5}l}
\hline\hline
&\multicolumn{2}{c}{(1)} 
&\multicolumn{2}{c}{(2)} 
&\multicolumn{2}{c}{(3)} 
&\multicolumn{2}{c}{(4)} \\
&\multicolumn{2}{c}{\text{$\Delta \ln$ hourly wage}} 
&\multicolumn{2}{c}{$\Delta \ln$ monthly wage} 
&\multicolumn{2}{c}{\text{$\Delta \ln$ monthly}} 
&\multicolumn{2}{c}{Working in same }\\
& & & & & \multicolumn{2}{c}{contractual hours} 
&\multicolumn{2}{c}{plant in $t+3$ (dummy) }
\\
\hline
\multicolumn{2}{l}{\textbf{Gross hourly wage  \euro{ 10.45 -- 11.99}}} \\
\quad $\times$ April -- January      &  0.0085^{***} & (0.0017)&  0.0052       & (0.0040)& -0.0033       &(0.0041) & -0.0233^{***} & (0.0045) \\
\quad $\times$ May -- February       &  0.0080^{***} & (0.0015)&  0.0088^{*}   & (0.0041)& 0.0009        &(0.0041) & -0.0206^{***} & (0.0045) \\
\quad $\times$ June -- March         &  0.0064^{***} & (0.0015)&  -0.0028      & (0.0041)& -0.0092^{*}   &(0.0039) & -0.0204^{***} & (0.0045) \\
\quad $\times$ July -- April         &  0.0118^{***} & (0.0014)&  0.0121^{**}  & (0.0038)& 0.0003        &(0.0038) & -0.0168^{***} & (0.0042) \\
\quad $\times$ August -- May         &  0.0118^{***} & (0.0017)&  0.0057       & (0.0039)& -0.0061       &(0.0039) & -0.0130^{**}  & (0.0043) \\
\quad $\times$ September -- June     &  0.0068^{***} & (0.0019)&  0.0035       & (0.0042)& -0.0033       &(0.0041) & -0.0148^{**}  & (0.0045) \\
\quad $\times$ October -- July       &  0.0633^{***} & (0.0014)&  0.0489^{***} & (0.0038)& -0.0144^{***} &(0.0037) & -0.0281^{***} & (0.0040) \\
\quad $\times$ November -- August    &  0.0589^{***} & (0.0022)&  0.0448^{***} & (0.0041)& -0.0141^{***} &(0.0037) & -0.0112^{**}  & (0.0043) \\
\quad $\times$ December -- September &  0.0617^{***} & (0.0031)&  0.0503^{***} & (0.0054)& -0.0114^{*}   &(0.0045) & -0.0235^{***} & (0.0045) \\[1.5ex]

\multicolumn{2}{l}{\textbf{Gross hourly wage  \euro{ 12.00 -- 12.99}}}&  \multicolumn{6}{c}{base} & \\[1.5ex]

\multicolumn{2}{l}{\textbf{Gross hourly wage \euro{ 13.00 -- 14.99}}} \\
\quad $\times$ April--January       &-0.0041^{** } & (0.0014)  & -0.0090^{*  } &  (0.0035)  &  -0.0049^{   }  &  (0.0036) &  0.0187^{***}  &  (0.0042)   \\
\quad $\times$ May--February        &-0.0069^{***} & (0.0014)  & -0.0093^{** } &  (0.0033)  &  -0.0024^{   }  &  (0.0032) &  0.0161^{***}  &  (0.0043)   \\
\quad $\times$ June--March          &-0.0028^{   } & (0.0015)  &  0.0002^{   } &  (0.0037)  &   0.0031^{   }  &  (0.0036) &  0.0197^{***}  &  (0.0042)   \\
\quad $\times$ July-April           &-0.0023^{   } & (0.0013)  & -0.0048^{   } &  (0.0032)  &  -0.0024^{   }  &  (0.0032) &  0.0182^{***}  &  (0.0039)   \\
\quad $\times$ August--May          &-0.0082^{***} & (0.0015)  & -0.0044^{   } &  (0.0035)  &   0.0038^{   }  &  (0.0033) &  0.0297^{***}  &  (0.0039)   \\
\quad $\times$ September--June      &-0.0120^{***} & (0.0017)  & -0.0107^{** } &  (0.0039)  &   0.0013^{   }  &  (0.0037) &  0.0255^{***}  &  (0.0042)   \\
\quad $\times$ October--July        &-0.0169^{***} & (0.0013)  & -0.0139^{***} &  (0.0034)  &   0.0031^{   }  &  (0.0033) &  0.0221^{***}  &  (0.0038)   \\
\quad $\times$ November--August     &-0.0165^{***} & (0.0021)  & -0.0155^{***} &  (0.0036)  &   0.0011^{   }  &  (0.0030) &  0.0391^{***}  &  (0.0041)   \\
\quad $\times$ December--September  &-0.0204^{***} & (0.0030)  & -0.0235^{***} &  (0.0053)  &  -0.0031^{   }  &  (0.0045) &  0.0312^{***}  &  (0.0043)   \\[1.5ex]

\multicolumn{2}{l}{\textbf{Gross hourly wage \euro{ 15.00 -- 20.99}}} \\
\quad $\times$ April -- January     &-0.0176^{***} & (0.0013)  & -0.0212^{***} &  (0.0033)  &  -0.0036^{   }  &  (0.0034) &  0.0438^{***}  &  (0.0037)   \\
\quad $\times$ May -- February      &-0.0267^{***} & (0.0013)  & -0.0220^{***} &  (0.0031)  &   0.0048^{   }  &  (0.0031) &  0.0454^{***}  &  (0.0037)   \\
\quad $\times$ June -- March        &-0.0137^{***} & (0.0012)  & -0.0131^{***} &  (0.0034)  &   0.0006^{   }  &  (0.0032) &  0.0531^{***}  &  (0.0037)   \\
\quad $\times$ July -- April        &-0.0152^{***} & (0.0010)  & -0.0138^{***} &  (0.0031)  &   0.0014^{   }  &  (0.0030) &  0.0559^{***}  &  (0.0034)   \\
\quad $\times$ August -- May        &-0.0200^{***} & (0.0013)  & -0.0116^{***} &  (0.0033)  &   0.0083^{** }  &  (0.0031) &  0.0663^{***}  &  (0.0035)   \\
\quad $\times$ September -- June    &-0.0230^{***} & (0.0016)  & -0.0216^{***} &  (0.0035)  &   0.0014^{   }  &  (0.0034) &  0.0645^{***}  &  (0.0036)   \\
\quad $\times$ October -- July      &-0.0352^{***} & (0.0012)  & -0.0289^{***} &  (0.0031)  &   0.0064^{*  }  &  (0.0030) &  0.0624^{***}  &  (0.0033)   \\
\quad $\times$ November -- August   &-0.0361^{***} & (0.0019)  & -0.0351^{***} &  (0.0032)  &   0.0010^{   }  &  (0.0027) &  0.0758^{***}  &  (0.0036)   \\
\quad $\times$ December -- September&-0.0349^{***} & (0.0028)  & -0.0312^{***} &  (0.0046)  &   0.0037^{   }  &  (0.0038) &  0.0711^{***}  &  (0.0037)   \\[1.5ex]

\multicolumn{2}{l}{\textbf{Gross hourly wage $\mathbf{\geqslant}$ \euro{21}}} \\
\quad $\times$ April -- January     &-0.0434^{***} & (0.0013)  & -0.0403^{***} &  (0.0032)  &   0.0031^{   }  &  (0.0034) &  0.0777^{***}  &  (0.0036)   \\
\quad $\times$ May -- February      &-0.0734^{***} & (0.0013)  & -0.0556^{***} &  (0.0030)  &   0.0178^{***}  &  (0.0030) &  0.0814^{***}  &  (0.0036)   \\
\quad $\times$ June -- March        &-0.0358^{***} & (0.0012)  & -0.0354^{***} &  (0.0033)  &   0.0004^{   }  &  (0.0032) &  0.0765^{***}  &  (0.0036)   \\
\quad $\times$ July -- April        &-0.0383^{***} & (0.0011)  & -0.0285^{***} &  (0.0030)  &   0.0098^{***}  &  (0.0030) &  0.0823^{***}  &  (0.0033)   \\
\quad $\times$ August -- May        &-0.0471^{***} & (0.0013)  & -0.0273^{***} &  (0.0032)  &   0.0199^{***}  &  (0.0030) &  0.0915^{***}  &  (0.0034)   \\
\quad $\times$ September -- June    &-0.0428^{***} & (0.0016)  & -0.0354^{***} &  (0.0034)  &   0.0074^{*  }  &  (0.0033) &  0.0965^{***}  &  (0.0035)   \\
\quad $\times$ October -- July      &-0.0595^{***} & (0.0012)  & -0.0454^{***} &  (0.0030)  &   0.0141^{***}  &  (0.0029) &  0.0965^{***}  &  (0.0032)   \\
\quad $\times$ November -- August   &-0.0617^{***} & (0.0019)  & -0.0570^{***} &  (0.0031)  &   0.0048^{   }  &  (0.0027) &  0.1062^{***}  &  (0.0035)   \\
\quad $\times$ December -- September&-0.0567^{***} & (0.0028)  & -0.0435^{***} &  (0.0046)  &   0.0133^{***}  &  (0.0037) &  0.1024^{***}  &  (0.0036)   \\
       
\hline
Employees
&\multicolumn{2}{c}{9,166,856}
&\multicolumn{2}{c}{9,166,856}
&\multicolumn{2}{c}{9,166,856}
&\multicolumn{2}{c}{10,412,969} \\
Observations 
& \multicolumn{2}{c}{65,609,448}
& \multicolumn{2}{c}{65,609,448}
& \multicolumn{2}{c}{65,609,448}
& \multicolumn{2}{c}{72,881,577}\\
\hline\hline
\end{tabular}
\begin{small}
\textit{Notes:} Columns (1)--(3) report regression coefficients from OLS estimations of equation (1). Column (4) reports regression coefficients from OLS estimation of equation (2). In specification (4), the dependent variable is equal to 1 if employee works in the same plant in $t+3$ as in $t$. All regressions are weighted. Standard errors in parentheses are clustered at the individual level. Asterisks indicate significance levels. *~$p<0.05$, **~$p<0.01$, ***~$p<0.001$. Sample size is larger for specification (4) since observations are only included in specifications (1)--(3) if employee works in the same establishment in $t$ and $t+3$. \\\textit{Data:} SE, 2022, monthly data, public sector excluded. 
\end{small}
\end{threeparttable}
\end{footnotesize}
\end{table}

\clearpage
\setcounter{figure}{0}
\setcounter{table}{0}
\renewcommand*{\thefigure}{\thesection\arabic{figure}}
\renewcommand*{\thetable}{\thesection\arabic{table}}
\section{Robustness}
\label{app:robustness}

\subsection{Time-constant wage groups from June wages}
\label{app:fixed_groups}

\begin{figure}[ht!]
\centering
\caption{Predictions based on regressions that include time-constant wage groups and 4-month differences}\label{fig:rob_fixed_groups}
\includegraphics[width=0.98\textwidth]{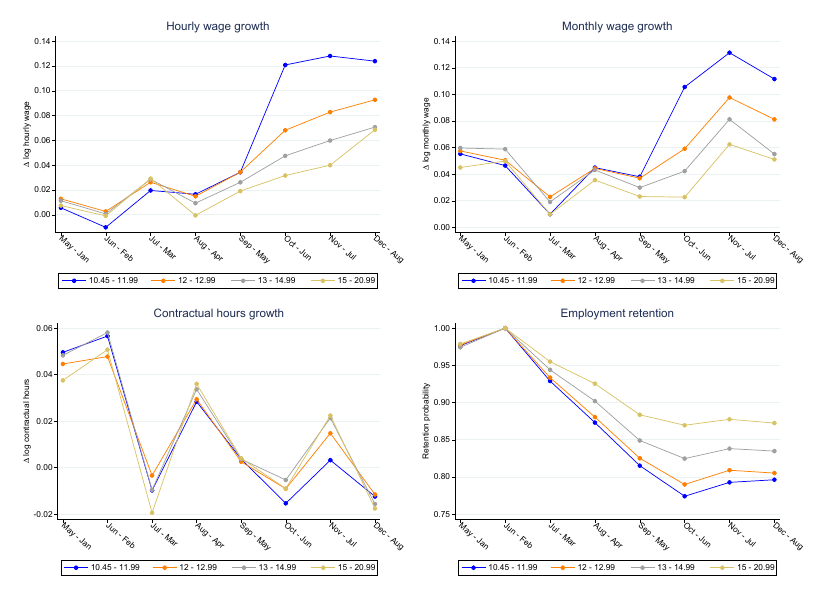}
\begin{justify}
\begin{small}{\textit{Notes:}
Each point is a prediction using estimates from equation~(\ref{eq:didid}) for the respective group $g$ and month $m$, i.e., each point is an estimate of $E[\ln y_{g,m+4} - \ln y_{g,m}]$. Regarding employment retention each point is a prediction using the estimates of a covariate-amended version of equation~(\ref{eq:did_Ret}) for the respective group $g$ and month $m$, i.e., each point is an estimate of $E[\ln y_{g,m+4}]$. The group $g$ of each individuals is fixed over time according to the respective individual's wage as of June. Hence the sample is also restricted to jobs that report a wage in June.\\
\textit{Source:} Earnings Survey, monthly panel January-December 2022, public sector excluded. }
\end{small}
\end{justify}
\end{figure}

\clearpage
\subsection{Inclusion of covariates}
\label{app:robustness_cov}

\begin{figure}[ht!]
\centering
\caption{Predictions based on regressions with covariates}\label{fig:rob_cov}
\includegraphics[width=0.98\textwidth]{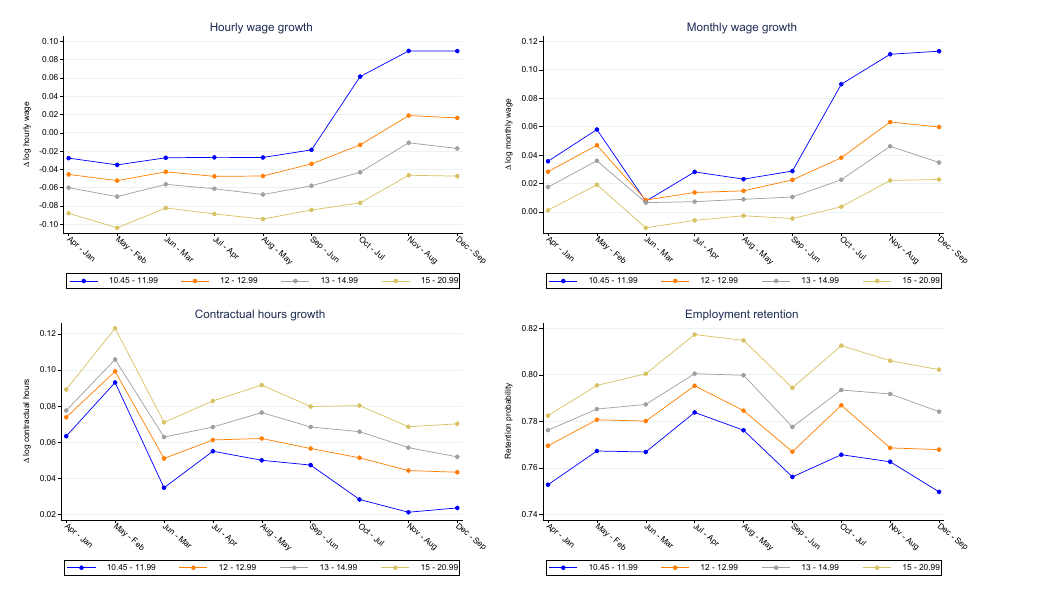}
\begin{justify}
\begin{small}{\textit{Notes:}
While covariates are included in the regressions, they are set equal to zero when obtaining the predictions.
Regarding hourly wage growth, monthly wage growth and contractual hours growth, each point is a prediction using the estimates of a covariate-amended version of equation~(\ref{eq:didid}) for the respective group $g$ and month $m$, i.e., each point is an estimate of $E[\ln y_{g,m+3} - \ln y_{g,m}]$. Regarding employment retention each point is a prediction using the estimates of a covariate-amended version of equation~(\ref{eq:did_Ret}) for the respective group $g$ and month $m$, i.e., each point is an estimate of $E[\ln y_{g,m+3}]$. 
The respective regressions include the following covariates:
gender, age, education (7 categories), foreign, job task (4 categories), part-time, minijobber, job tenure (9 categories), sector (2 digit), bargaining agreement, company size (7 categories), and federal state.
\\
\textit{Source:} Earnings Survey, monthly panel January-December 2022, public sector excluded. }
\end{small}
\end{justify}
\end{figure}

\newpage

\subsection{Heterogeneous effects by educational attainment}
\label{app:robustness_education}

\begin{figure}[ht!]
    \centering
    \caption{Hourly and monthly wage growth by education}
    \begin{subfigure}[b]{\textwidth}
        \centering
      \caption{Hourly wage growth}
        \includegraphics[width=0.9\textwidth]{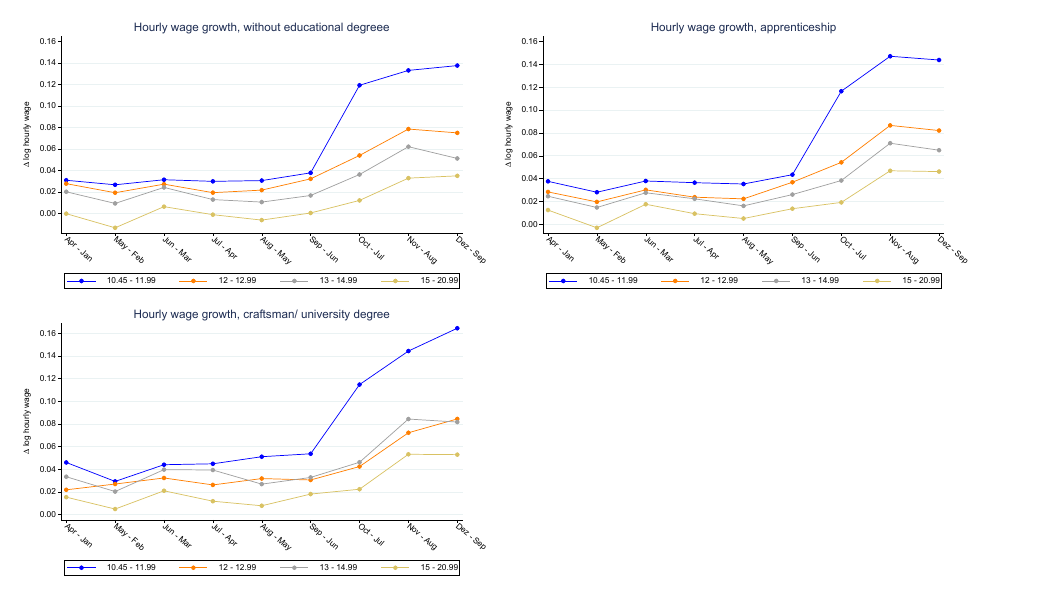}
            \end{subfigure}
    
   \vspace{0.5cm} % vertikaler Abstand zwischen den Grafiken

    \begin{subfigure}[b]{\textwidth}
        \centering
       \caption{Monthly wage growth}
                \includegraphics[width=0.9\textwidth]{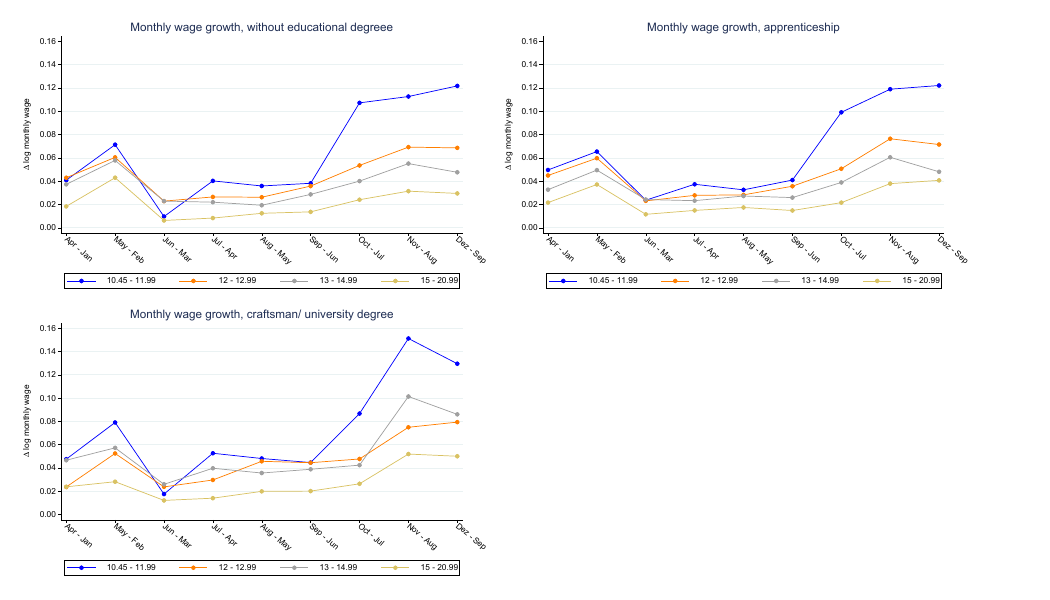}
            \end{subfigure}
\begin{justify}
\begin{small}{\textit{Notes:} Separate estimations and predictions by educational attainment. 
The upper part presents the development of hourly wage growth and the lower part presents the development of monthly wage growth. Each point is a prediction using the estimates of equation~(\ref{eq:didid}) for the respective group $g$ and month $m$, i.e., each point is an estimate of $E[\ln y_{g,m+3} - \ln y_{g,m}]$.  \\
\textit{Source:} Earnings Survey, monthly panel January-December 2022, public sector excluded. }
\end{small}
\end{justify}
\end{figure}

\begin{figure}[ht!]
    \centering
    \caption{Hours growth and employment retention by eduction}
    \begin{subfigure}[b]{\textwidth}
        \centering
      \caption{Contractual hours growth}
        \includegraphics[width=0.9\textwidth]{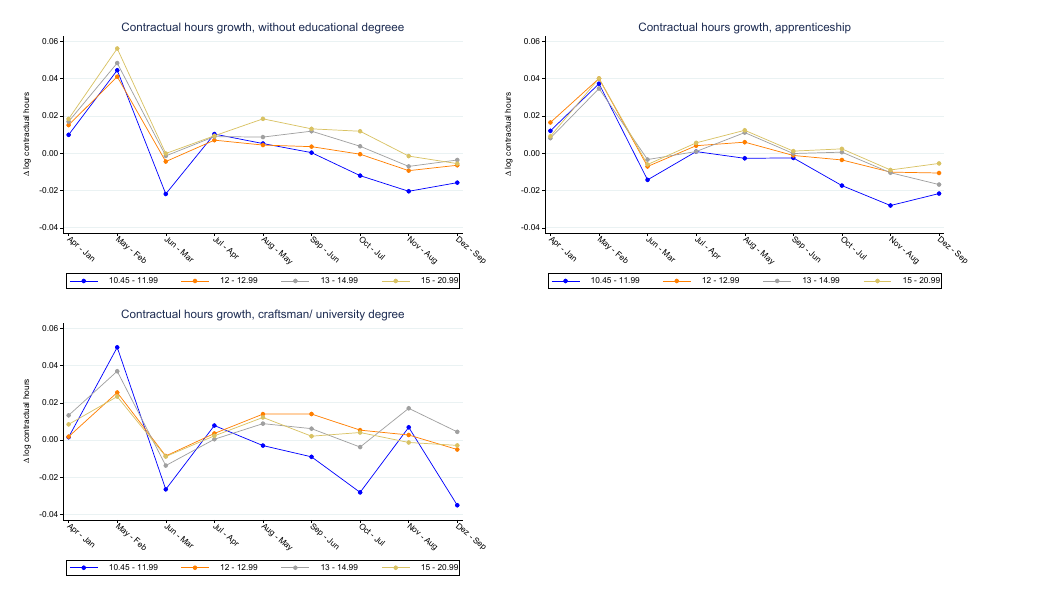}
            \end{subfigure}
    
    \vspace{0.5cm} % vertikaler Abstand zwischen den Grafiken

    \begin{subfigure}[b]{\textwidth}
        \centering
       \caption{Employment retention}
                \includegraphics[width=0.9\textwidth]{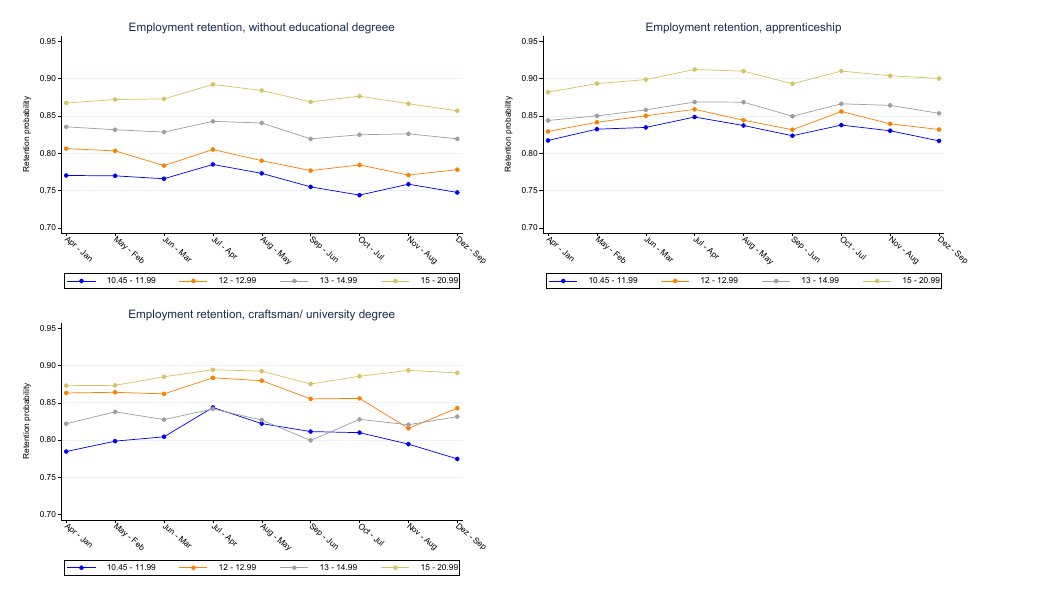}
            \end{subfigure}
\begin{justify}
\begin{small}{\textit{Notes:} Separate estimations and predictions by educational attainment. 
The upper part presents the development of monthly working hours growth and the lower part presents the development of employment retention. For monthly working hours, each point is a prediction using the estimates of equation~(\ref{eq:didid}) for the respective group $g$ and month $m$, i.e., each point is an estimate of $E[\ln y_{g,m+3} - \ln y_{g,m}]$. For employment retention, each point is a prediction using estimates of equation~(\ref{eq:did_Ret}) for the respective group $g$ and month $m$, i.e., each point is an estimate of $E[\ln E_{g,m+3}]$.
\textit{Source:} Earnings Survey, monthly panel January-December 2022, public sector excluded. }
\end{small}
\end{justify}
\end{figure}

\clearpage

\subsection{Heterogeneous effects by tenure}
\label{app:robustness_tenure}
\begin{figure}[ht!]
    \centering
    \caption{Hourly and monthly wage growth by tenure}
    \begin{subfigure}[b]{\textwidth}
        \centering
      \caption{Hourly wage growth}
        \includegraphics[width=0.9\textwidth]{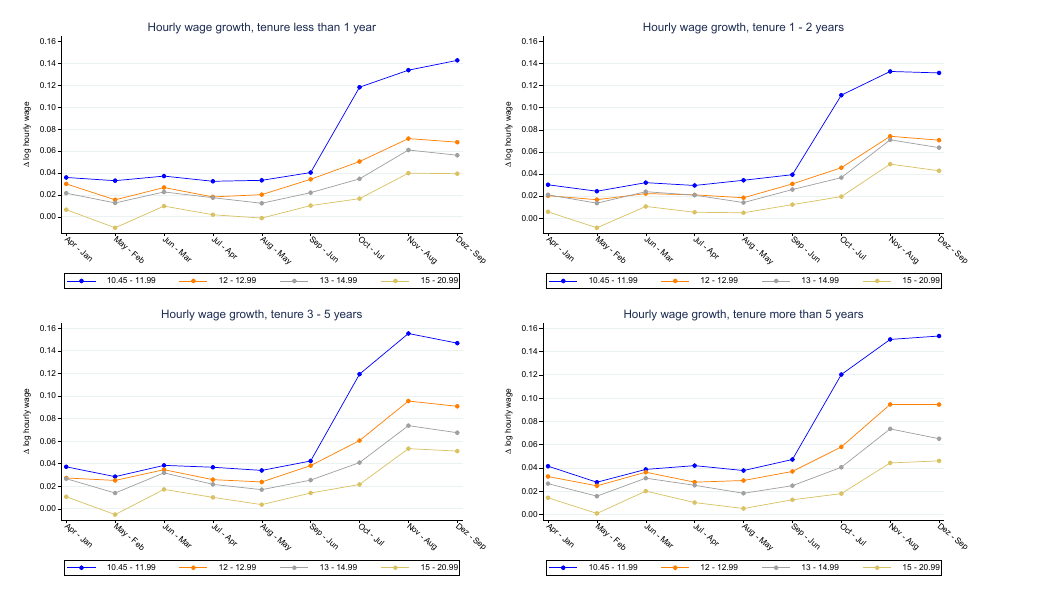}
            \end{subfigure}
    
    \vspace{0.5cm} % vertikaler Abstand zwischen den Grafiken
   
    \begin{subfigure}[b]{\textwidth}
        \centering
       \caption{Monthly wage growth}
                \includegraphics[width=0.9\textwidth]{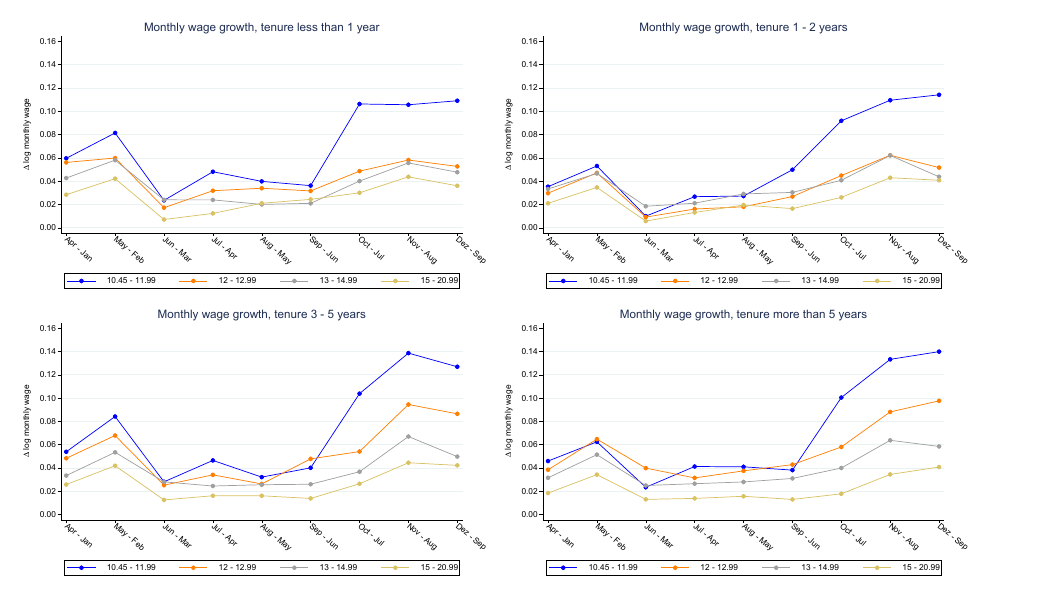}
            \end{subfigure}
\begin{justify}
\begin{small}{\textit{Notes:} Separate estimations and predictions by tenure group. 
The upper part presents the development of hourly wage growth and the lower part presents the development of monthly wage growth. Each point is a prediction using the estimates of equation~(\ref{eq:didid}) for the respective group $g$ and month $m$, i.e., each point is an estimate of $E[\ln y_{g,m+3} - \ln y_{g,m}]$.  \\
\textit{Source:} Earnings Survey, monthly panel January-December 2022, public sector excluded. }
\end{small}
\end{justify}
\end{figure}

\begin{figure}[ht!]
    \centering
    \caption{Hours growth and employment retention by tenure}
    \begin{subfigure}[b]{\textwidth}
        \centering
      \caption{Contractual hours growth}
        \includegraphics[width=0.9\textwidth]{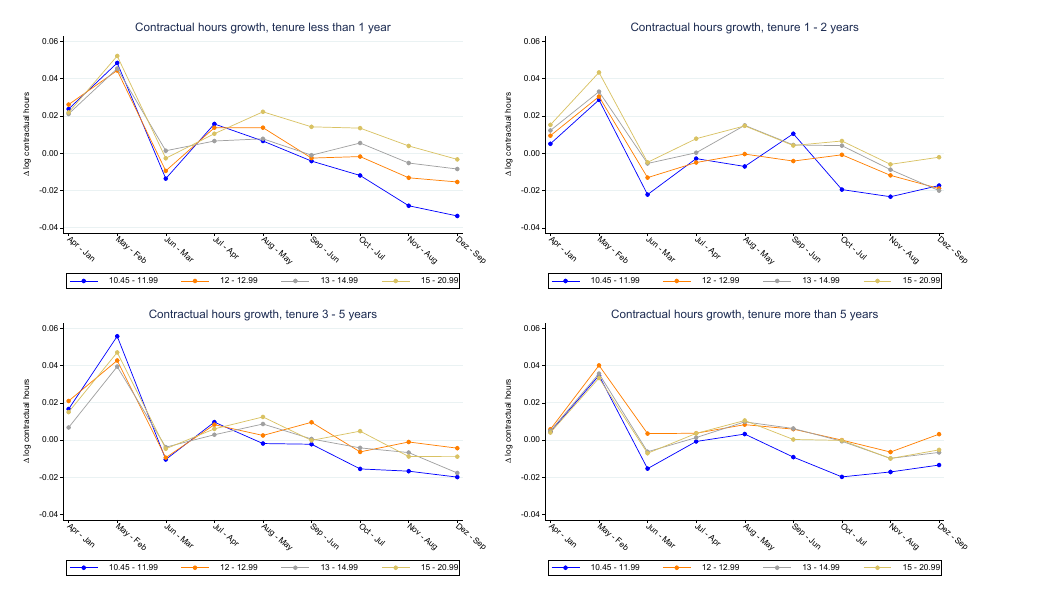}
            \end{subfigure}
    
    \vspace{0.5cm} % vertikaler Abstand zwischen den Grafiken

    \begin{subfigure}[b]{\textwidth}
        \centering
       \caption{Employment retention}
                \includegraphics[width=0.9\textwidth]{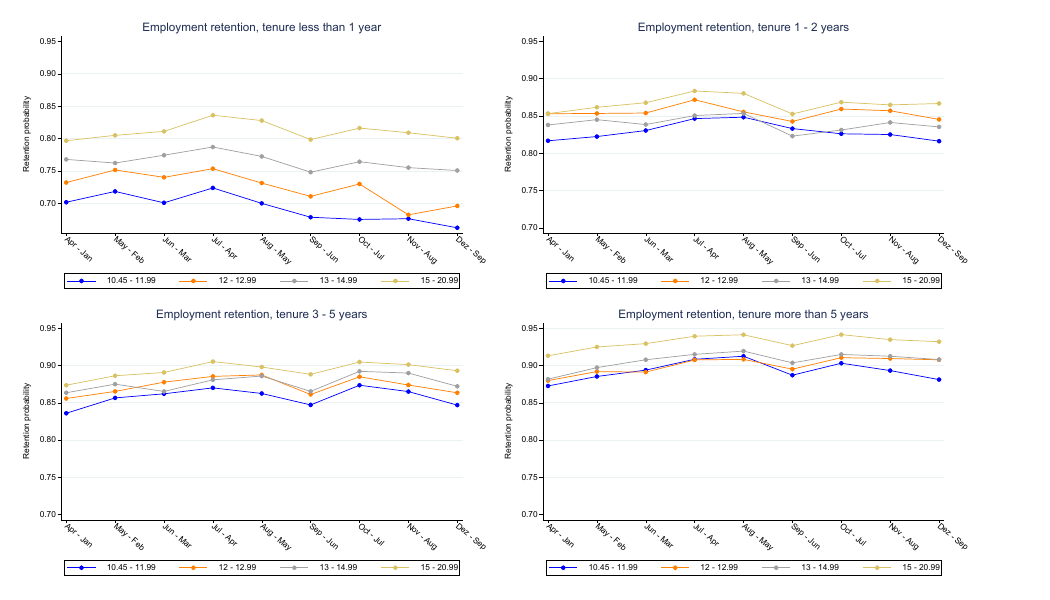}
            \end{subfigure}
\begin{justify}
\begin{small}{\textit{Notes:} Separate estimations and predictions by tenure group. 
The upper part presents the development of monthly working hours growth and the lower part presents the development of employment retention. For monthly working hours, each point is a prediction using the estimates of equation~(\ref{eq:didid}) for the respective group $g$ and month $m$, i.e., each point is an estimate of $E[\ln y_{g,m+3} - \ln y_{g,m}]$. For employment retention, each point is a prediction using estimates of equation~(\ref{eq:did_Ret}) for the respective group $g$ and month $m$, i.e., each point is an estimate of $E[\ln E_{g,m+3}]$.
\textit{Source:} Earnings Survey, monthly panel January-December 2022, public sector excluded. }
\end{small}
\end{justify}
\end{figure}

\clearpage

\subsection{Heterogeneous effects by establishment size}
\label{app:robustness_size}

\begin{figure}[ht!]
    \centering
    \caption{Hourly and monthly wage growth by establishment size}
    \begin{subfigure}[b]{\textwidth}
        \centering
      \caption{Hourly wage growth}
        \includegraphics[width=0.9\textwidth]{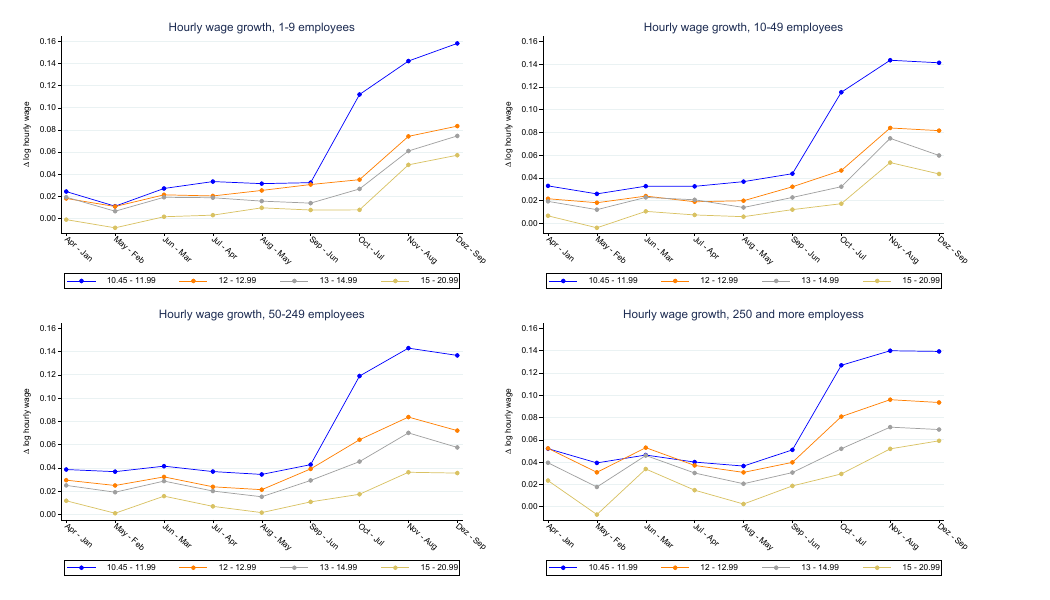}
            \end{subfigure}
    
    \vspace{0.5cm} % vertikaler Abstand zwischen den Grafiken

    \begin{subfigure}[b]{\textwidth}
        \centering
       \caption{Monthly wage growth}
                \includegraphics[width=0.9\textwidth]{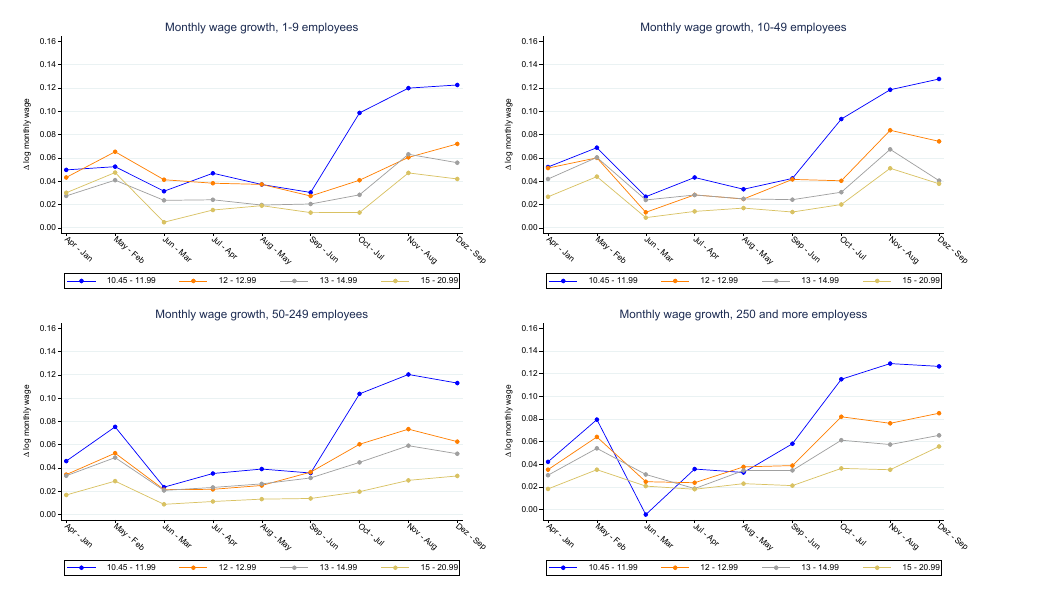}
            \end{subfigure}
\begin{justify}
\begin{small}{\textit{Notes:} Separate estimations and predictions by establishment size. 
The upper part presents the development of hourly wage growth and the lower part presents the development of monthly wage growth. Each point is a prediction using the estimates of equation~(\ref{eq:didid}) for the respective group $g$ and month $m$, i.e., each point is an estimate of $E[\ln y_{g,m+3} - \ln y_{g,m}]$.  \\
\textit{Source:} Earnings Survey, monthly panel January-December 2022, public sector excluded. }
\end{small}
\end{justify}
\end{figure}

\begin{figure}[ht!]
    \centering
    \caption{Hours growth and employment retention by establishment size}
    \begin{subfigure}[b]{\textwidth}
        \centering
      \caption{Contractual hours growth}
        \includegraphics[width=0.9\textwidth]{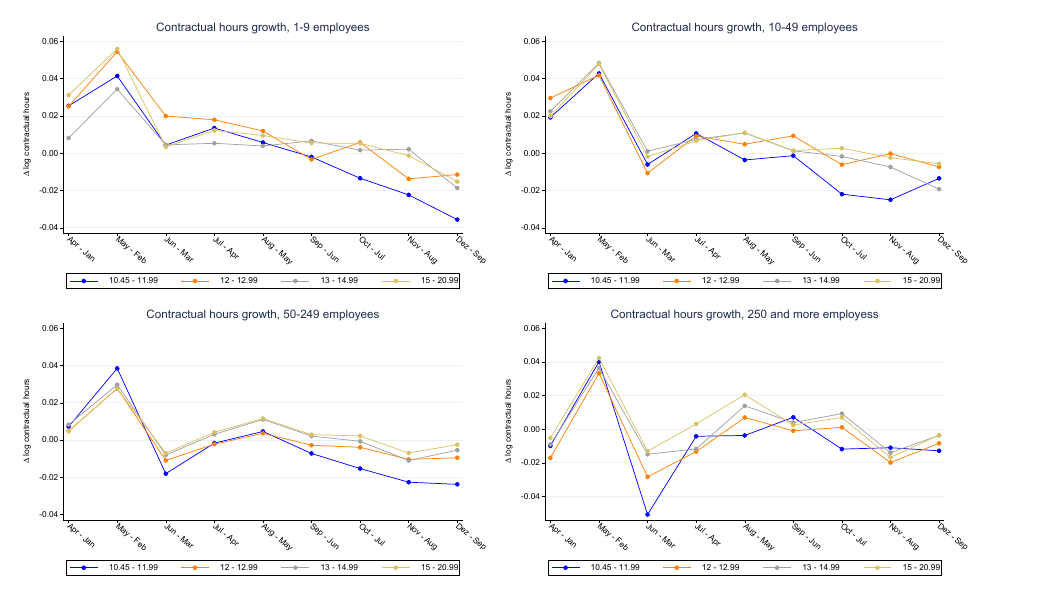}
            \end{subfigure}
    
    \vspace{0.5cm} % vertikaler Abstand zwischen den Grafiken

    \begin{subfigure}[b]{\textwidth}
        \centering
       \caption{Employment retention}
                \includegraphics[width=0.9\textwidth]{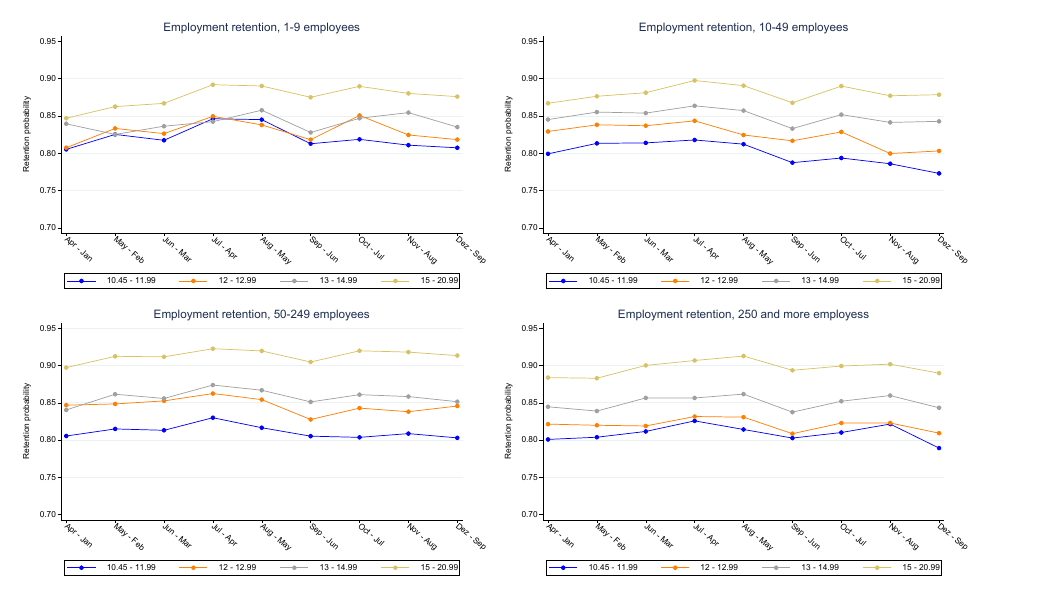}
            \end{subfigure}
\begin{justify}
\begin{small}{\textit{Notes:} Separate estimations and predictions by establishment size. 
The upper part presents the development of monthly working hours growth and the lower part presents the development of employment retention. For monthly working hours, each point is a prediction using the estimates of equation~(\ref{eq:didid}) for the respective group $g$ and month $m$, i.e., each point is an estimate of $E[\ln y_{g,m+3} - \ln y_{g,m}]$. For employment retention, each point is a prediction using estimates of equation~(\ref{eq:did_Ret}) for the respective group $g$ and month $m$, i.e., each point is an estimate of $E[\ln E_{g,m+3}]$.
\textit{Source:} Earnings Survey, monthly panel January-December 2022, public sector excluded. }
\end{small}
\end{justify}
\end{figure}

\clearpage

\subsection{Worker fixed effects regressions}
\label{app:robustness_FE}

\begin{figure}[ht!]
\centering
\caption{Predictions based on regressions with worker fixed effects}\label{fig:rob_FE}
\includegraphics[width=0.98\textwidth]{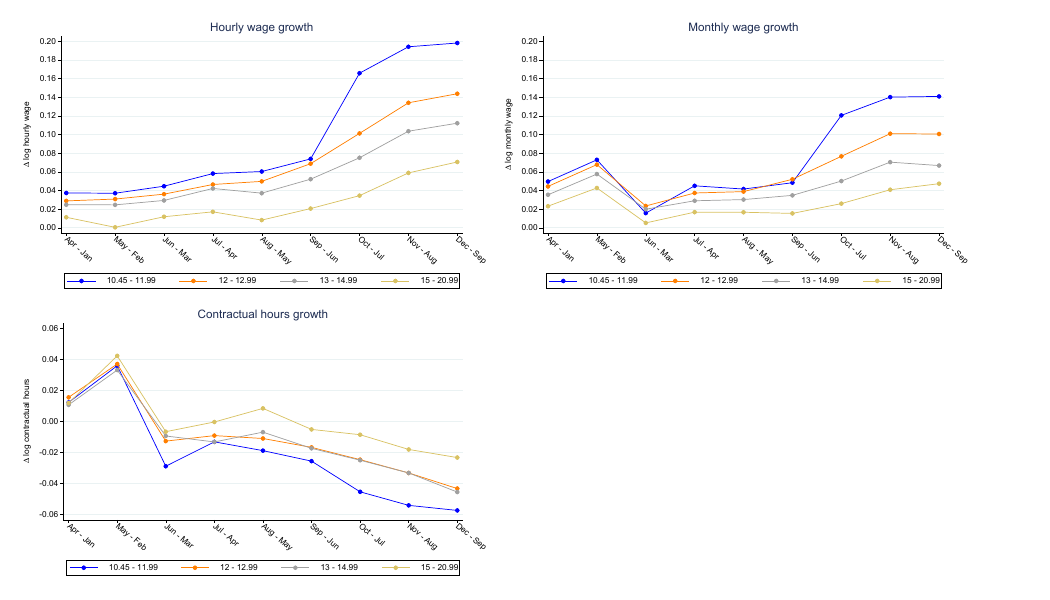}
\begin{justify}
\begin{small}{\textit{Notes:}
For each of the three outcome variables, each point is a prediction using the estimates for the respective group $g$ and month $m$  of a worker fixed effects amended version of equation~(\ref{eq:didid}), i.e., each point is an estimate of $E[\ln y_{g,m+3} - \ln y_{g,m}]$. 
Note that in FE regressions, the level of the four lines is identified by employees moving into and out of the groups. Therefore, in all three graphs, each of the four lines have been shifted parallel by the respective January value obtained from the baseline regressions (displayed in Figures~\ref{fig:12_base_wage} and \ref{fig:12_base_employment}). Since this has no impact on the development of the four lines, it does not affect the difference-in-differences treatment effect.\\
\textit{Source:} Earnings Survey, monthly panel January-December 2022, public sector excluded. }
\end{small}
\end{justify}
\end{figure}

\clearpage

\subsection{Growth rates over one month}
\label{app:robustness_1month}

\begin{figure}[ht!]
\centering
\caption{One month differences}\label{fig:rob_1month}
\includegraphics[width=0.98\textwidth]{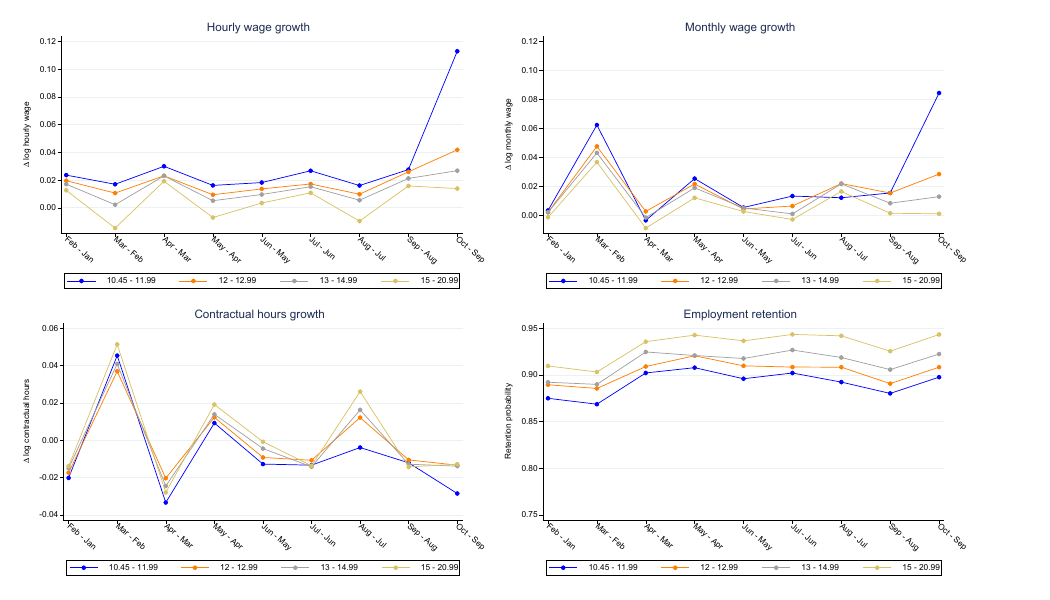}
\begin{justify}
\begin{small}{\textit{Notes:}
For each of the three outcome variables, each point is a prediction using the estimates for the respective group $g$ and month $m$  of an adjusted one-month version of equation~(\ref{eq:didid}), i.e., each point is an estimate of $E[\ln y_{g,m+1} - \ln y_{g,m}]$. \\
\textit{Source:} Earnings Survey, monthly panel January-October 2022, public sector excluded. }
\end{small}
\end{justify}
\end{figure}

\clearpage
\setcounter{figure}{0}
\setcounter{table}{0}
\renewcommand*{\thefigure}{\thesection\arabic{figure}}
\renewcommand*{\thetable}{\thesection\arabic{table}}
\section{Alternative specifications regarding aggregate employment effects}
\label{app:employment}

\begin{figure}[ht!]
\centering
\caption{County-level bite of the \euro{12}-minimum wage increase. }\label{fig:bite}
\includegraphics[width=0.5\textwidth,height=0.5\textheight]{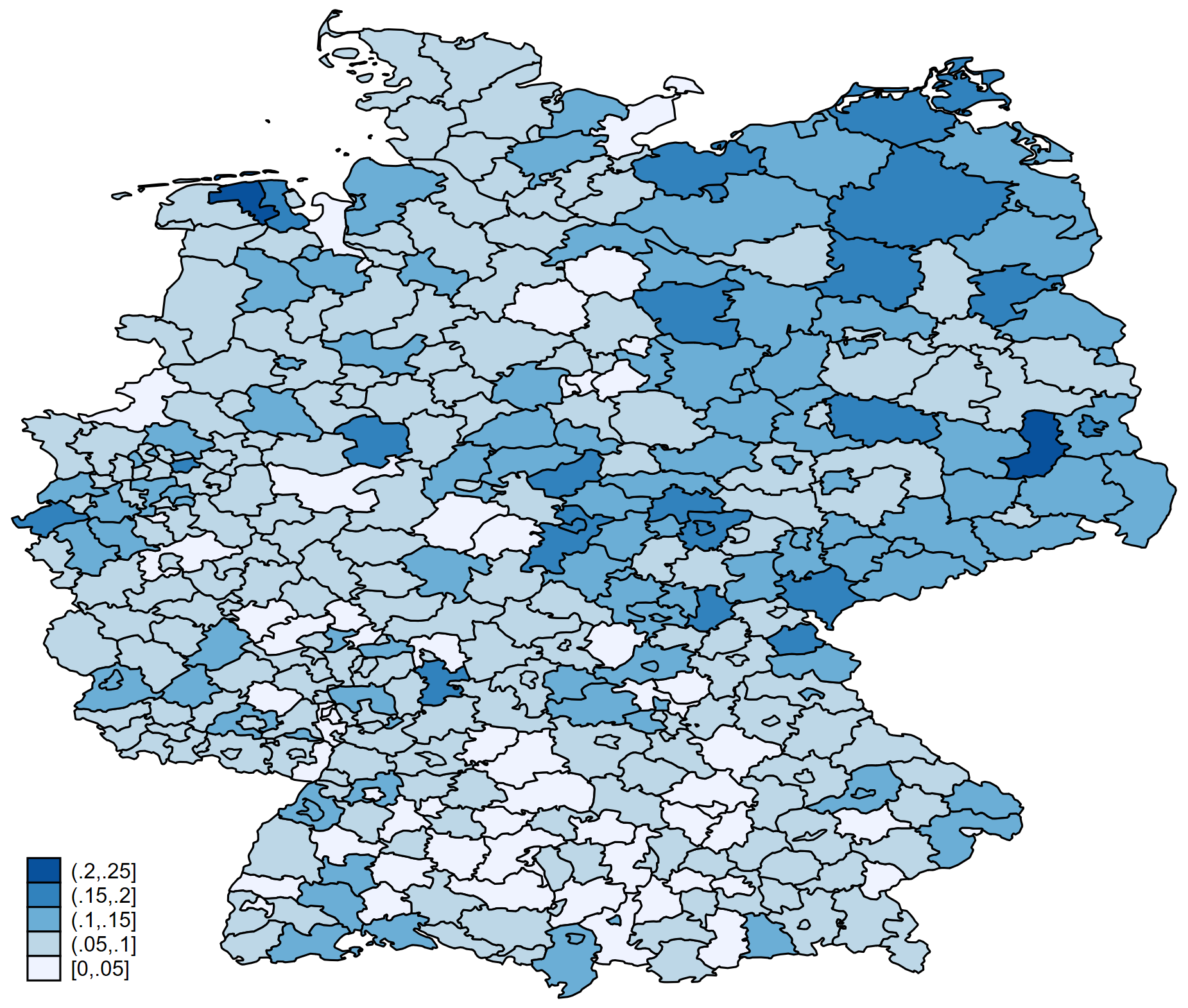}
\begin{justify}
\begin{small}{%\textit{Notes:} 
\textit{Source:} Earnings Survey, April 2022, public sector excluded. }
\end{small}
\end{justify}
\end{figure}

\begin{table}[ht!]
\begin{center}
\caption{Region-level employment effects, full regression table}
\label{tab:employment_full_table}
\resizebox{1.0\linewidth}{!}{
\begin{threeparttable}
\begin{tabular}{lC{1.6cm}C{1.8cm}C{1.6cm}C{1.8cm}C{1.6cm}C{1.8cm}}
\hline\hline
    &\multicolumn{2}{c}{Total Employment}&\multicolumn{2}{c}{Minijobs}&\multicolumn{2}{c}{Regular Employment} \\
    \cmidrule(lr){2-3}\cmidrule(lr){4-5}\cmidrule(lr){6-7}
    &\multicolumn{1}{c}{(1)}&\multicolumn{1}{c}{(2)}&\multicolumn{1}{c}{(3)}&\multicolumn{1}{c}{(4)}&\multicolumn{1}{c}{(5)}&\multicolumn{1}{c}{(6)}\\
    & \makecell{log} &\makecell{12 months \\log \\difference} & \makecell{log} & \makecell{12 months \\log \\difference} & \makecell{log} & \makecell{12 months \\log \\difference} \\
\hline
$Bite \times January$ & base & base & base & base & base & base \\[0.3ex]
$Bite \times February$ & -0.009*** &       0.005         &      -0.025*** &      -0.009         &      -0.010*** &       0.009*  \\
    &     (0.003)         &     (0.006)         &     (0.008)         &     (0.012)         &     (0.003)         &     (0.005)         \\[.3ex]
$Bite \times March$ &      -0.008         &       0.004         &      -0.053*** &      -0.040** &      -0.001         &       0.020*  \\
                    &     (0.011)         &     (0.011)         &     (0.011)         &     (0.018)         &     (0.012)         &     (0.011)         \\[.3ex]
$Bite \times April$ &       0.004         &       0.023         &      -0.032*  &       0.002         &       0.012         &       0.031** \\
                    &     (0.019)         &     (0.017)         &     (0.018)         &     (0.028)         &     (0.019)         &     (0.015)         \\[.3ex]
$Bite \times May$ &      -0.001         &       0.026         &      -0.045*  &       0.014         &       0.008         &       0.030** \\
                    &     (0.022)         &     (0.016)         &     (0.024)         &     (0.030)         &     (0.022)         &     (0.015)         \\[.3ex]
$Bite \times June$ &      -0.004         &       0.018         &      -0.033         &       0.009         &       0.011         &       0.023*  \\
                    &     (0.026)         &     (0.013)         &     (0.029)         &     (0.032)         &     (0.023)         &     (0.012)         \\[.3ex]
$Bite \times July$ &       0.002         &       0.009         &      -0.022         &      -0.028         &       0.003         &       0.013         \\
                    &     (0.030)         &     (0.014)         &     (0.043)         &     (0.039)         &     (0.025)         &     (0.013)         \\[.3ex]
$Bite \times August$ &       0.013         &       0.004         &      -0.008         &      -0.024         &       0.001         &       0.010         \\
                    &     (0.034)         &     (0.020)         &     (0.050)         &     (0.045)         &     (0.030)         &     (0.020)         \\[.3ex]
$Bite \times September$ &      -0.020         &      -0.000         &      -0.028         &       0.003         &      -0.011         &       0.005         \\
                    &     (0.032)         &     (0.022)         &     (0.048)         &     (0.050)         &     (0.030)         &     (0.022)         \\[.3ex]
$Bite \times October$ &      -0.033         &      -0.007         &      -0.083*  &       0.008         &      -0.030         &       0.001         \\
                    &     (0.030)         &     (0.023)         &     (0.045)         &     (0.050)         &     (0.029)         &     (0.023)         \\[.3ex]
$Bite \times November$ &      -0.052** &      -0.007         &      -0.128*** &       0.004         &      -0.043*  &       0.006         \\
                    &     (0.025)         &     (0.023)         &     (0.043)         &     (0.052)         &     (0.025)         &     (0.023)         \\[.3ex]
$Bite \times December$ &      -0.074*** &      -0.012         &      -0.170*** &       0.004         &      -0.064*** &       0.003         \\
                    &     (0.023)         &     (0.023)         &     (0.043)         &     (0.054)         &     (0.023)         &     (0.023)         \\[.5ex]
$Bite$ &      -2.153** &      -0.098*** &      -4.698*** &      -0.438*** &      -1.648*  &      -0.045** \\
                    &     (0.897)         &     (0.017)         &     (0.951)         &     (0.034)         &     (0.912)         &     (0.018)         \\[.5ex]
\hline
\end{tabular}
\raggedleft{Continued overleaf}
\end{threeparttable}
}
\end{center}
\end{table}

\setcounter{table}{0}
\begin{table}[ht!]
\begin{center}
\caption{Cotinued}
\resizebox{1.0\linewidth}{!}{
\begin{threeparttable}
\begin{tabular}{lC{1.6cm}C{1.8cm}C{1.6cm}C{1.8cm}C{1.6cm}C{1.8cm}}
\hline\hline
    &\multicolumn{2}{c}{Total Employment}&\multicolumn{2}{c}{Minijobs}&\multicolumn{2}{c}{Regular Employment} \\
    \cmidrule(lr){2-3}\cmidrule(lr){4-5}\cmidrule(lr){6-7}
    &\multicolumn{1}{c}{(1)}&\multicolumn{1}{c}{(2)}&\multicolumn{1}{c}{(3)}&\multicolumn{1}{c}{(4)}&\multicolumn{1}{c}{(5)}&\multicolumn{1}{c}{(6)}\\
    & \makecell{log} &\makecell{12 months \\log \\difference} & \makecell{log} & \makecell{12 months \\log \\difference} & \makecell{log} & \makecell{12 months \\log \\difference} \\
\hline
$January$ & base & base & base & base & base & base \\[.3ex]
$February$  &       0.003*** &       0.002*** &       0.003*** &       0.004*** &       0.005*** &       0.001*** \\
                    &     (0.000)         &     (0.000)         &     (0.001)         &     (0.001)         &     (0.000)         &     (0.000)         \\[.3ex]
$March$ &       0.007***&       0.001         &       0.009***&       0.006***&       0.008***&      -0.000         \\
                    &     (0.001)         &     (0.001)         &     (0.001)         &     (0.001)         &     (0.001)         &     (0.001)         \\[.3ex]
$April$ &       0.009***&      -0.001         &       0.013***&       0.004*  &       0.008***&      -0.002         \\
                    &     (0.002)         &     (0.001)         &     (0.002)         &     (0.002)         &     (0.002)         &     (0.001)         \\[.3ex]
$May$ &       0.014***&      -0.001         &       0.023***&       0.001         &       0.011***&      -0.001         \\
                    &     (0.002)         &     (0.001)         &     (0.002)         &     (0.003)         &     (0.002)         &     (0.001)         \\[.3ex]
$June$ &       0.015***&      -0.005***&       0.026***&      -0.015***&       0.013***&      -0.002** \\
                    &     (0.002)         &     (0.001)         &     (0.003)         &     (0.003)         &     (0.002)         &     (0.001)         \\[.3ex]
$July$ &       0.012***&      -0.007***&       0.024***&      -0.022***&       0.014***&      -0.003*** \\
                    &     (0.003)         &     (0.001)         &     (0.004)         &     (0.003)         &     (0.002)         &     (0.001)         \\[.3ex]
$August$ &       0.016*** &      -0.008*** &       0.022*** &      -0.024*** &       0.015*** &      -0.004** \\
                    &     (0.003)         &     (0.002)         &     (0.004)         &     (0.004)         &     (0.002)         &     (0.002)         \\[.3ex]
$September$ &       0.026*** &      -0.008*** &       0.025*** &      -0.025*** &       0.020*** &      -0.004** \\
                    &     (0.003)         &     (0.002)         &     (0.004)         &     (0.004)         &     (0.002)         &     (0.002)         \\[.3ex]
$October$ &       0.026*** &      -0.008*** &       0.031*** &      -0.024*** &       0.021*** &      -0.005***\\
                    &     (0.002)         &     (0.002)         &     (0.004)         &     (0.004)         &     (0.002)         &     (0.002)         \\[.3ex]
$November$ & 0.026*** & -0.009*** &       0.036*** &      -0.022*** &       0.021*** &      -0.007*** \\
                    &     (0.002)         &     (0.002)         &     (0.003)         &     (0.004)         &     (0.002)         &     (0.002)         \\[.3ex]
$December$ & 0.021*** & -0.009*** & 0.033*** & -0.020*** & 0.016*** & -0.008*** \\
                    &     (0.002)         &     (0.002)         &     (0.003)         &     (0.004)         &     (0.002)         &     (0.002)         \\[.5ex]
Constant &      11.404*** &       0.025*** &       9.842*** &       0.040*** &      11.112*** &       0.024*** \\
                    &     (0.087)         &     (0.001)         &     (0.089)         &     (0.003)         &     (0.089)         &     (0.002)         \\[.5ex]
\hline
Cluster        &        400         &        400         &        400         &        400         &        400         &        400         \\
Observations        &        4800         &        4800         &        4800         &        4800         &        4800         &        4800         \\
\hline\hline
\end{tabular}
\begin{small} \textit{Notes:} Full table of coefficients from region-level difference-in-difference regressions as specified in equation (\ref{eq:region}). Outcome variables are the (unadjusted) logarithm and, respectively, the 12-month difference of the (adjusted) logarithm of total employment, minijobs, and regular social security employees. Standard errors in parentheses are clustered at the county level. Asterisks indicate significance levels *~$p<0.05$, **~$p<0.01$, ***~$p<0.001$. \textit{Source:} County-level administrative employment data of the Federal Employment Agency, and Earnings Survey of April 2022 for the regionally aggregated bite.
\end{small}
\end{threeparttable}
}
\end{center}
\end{table}

\begin{table}[ht!]
\begin{center}
\caption{Region-level employment effects, unadjusted}
\label{tab:employment_unadj}
\resizebox{1.0\linewidth}{!}{
\begin{threeparttable}
\begin{tabular}{lC{1.6cm}C{1.8cm}C{1.6cm}C{1.8cm}C{1.6cm}C{1.8cm}}
\hline\hline
    &\multicolumn{2}{c}{Total Employment}&\multicolumn{2}{c}{Minijobs}&\multicolumn{2}{c}{Regular Employment} \\
    \cmidrule(lr){2-3}\cmidrule(lr){4-5}\cmidrule(lr){6-7}
    &\multicolumn{1}{c}{(1)}&\multicolumn{1}{c}{(2)}&\multicolumn{1}{c}{(3)}&\multicolumn{1}{c}{(4)}&\multicolumn{1}{c}{(5)}&\multicolumn{1}{c}{(6)}\\
    & \makecell{log} &\makecell{12 months \\log \\difference} & \makecell{log} & \makecell{12 months \\log \\difference} & \makecell{log} & \makecell{12 months \\log \\difference} \\
\hline
$Bite \times January$ & base & base & base & base & base & base \\[.4ex]
$Bite \times February$ &      -0.006*  &       0.007         &       0.005         &       0.014         &      -0.010***&       0.007         \\
                    &     (0.003)         &     (0.006)         &     (0.008)         &     (0.012)         &     (0.003)         &     (0.005)         \\[.4ex]
$Bite \times March$ &      -0.001         &       0.009         &       0.007         &       0.006         &       0.000         &       0.017         \\
                    &     (0.011)         &     (0.011)         &     (0.011)         &     (0.018)         &     (0.012)         &     (0.011)         \\[.4ex]
$Bite \times April$ &       0.015         &       0.030*  &       0.057***&       0.071** &       0.013         &       0.026*  \\
                    &     (0.019)         &     (0.017)         &     (0.018)         &     (0.028)         &     (0.019)         &     (0.015)         \\[.4ex]
$Bite \times May$ &       0.013         &       0.034** &       0.074***&       0.106***&       0.010         &       0.024         \\
                    &     (0.022)         &     (0.016)         &     (0.024)         &     (0.030)         &     (0.022)         &     (0.015)         \\[.4ex]
$Bite \times June$ &       0.014         &       0.029** &       0.115***&       0.123***&       0.014         &       0.014         \\
                    &     (0.026)         &     (0.013)         &     (0.029)         &     (0.032)         &     (0.023)         &     (0.012)         \\[.4ex]
$Bite \times July$ &       0.023         &       0.022         &       0.155***&       0.110***&       0.006         &       0.003         \\
                    &     (0.030)         &     (0.014)         &     (0.043)         &     (0.039)         &     (0.025)         &     (0.013)         \\[.4ex]
$Bite \times August$ &       0.038         &       0.019         &       0.199***&       0.137***&       0.005         &      -0.002         \\
                    &     (0.034)         &     (0.020)         &     (0.050)         &     (0.045)         &     (0.030)         &     (0.020)         \\[.4ex]
$Bite \times September$ &       0.009         &       0.017         &       0.209***&       0.186***&      -0.007         &      -0.009         \\
                    &     (0.032)         &     (0.022)         &     (0.048)         &     (0.050)         &     (0.030)         &     (0.022)         \\[.4ex]
 $Bite \times October$ &      -0.001         &       0.012         &       0.183***&       0.214***&      -0.025         &      -0.014         \\
                    &     (0.030)         &     (0.023)         &     (0.045)         &     (0.050)         &     (0.029)         &     (0.023)         \\[.4ex]
 $Bite \times November$ &      -0.016         &       0.015         &       0.168***&       0.233***&      -0.038         &      -0.011         \\
                    &     (0.025)         &     (0.023)         &     (0.043)         &     (0.052)         &     (0.025)         &     (0.023)         \\[.4ex]
$Bite \times December$ &      -0.035         &       0.012         &       0.155***&       0.256***&      -0.059** &      -0.016         \\
                    &     (0.023)         &     (0.023)         &     (0.043)         &     (0.054)         &     (0.023)         &     (0.023)         \\
\hline
Cluster        &        400         &        400         &        400         &        400         &        400         &        400         \\
Observations        &        4800         &        4800         &        4800         &        4800         &        4800         &        4800         \\
\hline\hline
\end{tabular}
\begin{small} \textit{Notes:} Treatment interactions from region-level difference-in-difference regressions as specified in equation (\ref{eq:region}). Outcome variables are the (unadjusted) logarithm and, respectively, the 12-month difference of the (adjusted) logarithm of total employment, minijobs, and regular social security employees. Standard errors in parentheses are clustered at the county level. Asterisks indicate significance levels *~$p<0.05$, **~$p<0.01$, ***~$p<0.001$. \textit{Source:} County-level administrative employment data of the Federal Employment Agency, and Earnings Survey of April 2022 for the regionally aggregated bite.
\end{small}
\end{threeparttable}
}
\end{center}
\end{table}

\renewcommand{\arraystretch}{1.2}
\begin{table}[ht!]
\begin{center}
\caption{Unconditional region-level minijob development}
\label{tab:app_minijob}
\begin{threeparttable}
\begin{tabular}{L{3.5cm}R{2cm}L{2cm}}
\hline\hline
                    &\multicolumn{2}{c}{(1)}           \\
                    &\multicolumn{2}{c}{log Minijobs}    \\
\hline
$Year_{2021} \times January$ &      -0.028***&     (0.001)\\
$Year_{2021} \times February$ &      -0.030***&     (0.001)\\
$Year_{2021} \times March$ &      -0.025***&     (0.001)\\
$Year_{2021} \times April$ &      -0.020***&     (0.001)\\
$Year_{2021} \times May$ &      -0.010***&     (0.001)\\
$Year_{2021} \times June$ &       0.012***&     (0.001)\\
$Year_{2021} \times July$ &       0.022***&     (0.001)\\
$Year_{2021} \times August$ &       0.023***&     (0.001)\\
$Year_{2021} \times September$ &       0.024***&     (0.001)\\
$Year_{2021} \times October$ &       0.024***&     (0.001)\\
$Year_{2021} \times November$ &       0.025***&     (0.001)\\
$Year_{2021} \times December$ &       0.016***&     (0.000)\\[1ex]
$Year_{2022} \times Januar$  & base & \\
$Year_{2022} \times February$  &       0.003***&     (0.000)\\
$Year_{2022} \times March$  &       0.010***&     (0.000)\\
$Year_{2022} \times April$  &       0.018***&     (0.001)\\
$Year_{2022} \times May$  &       0.029***&     (0.001)\\
$Year_{2022} \times June$  &       0.036***&     (0.001)\\
$Year_{2022} \times July$  &       0.037***&     (0.001)\\
$Year_{2022} \times August$  &       0.039***&     (0.002)\\
$Year_{2022} \times September$  &       0.043***&     (0.001)\\
$Year_{2022} \times October$  &       0.047***&     (0.001)\\
$Year_{2022} \times November$  &       0.050***&     (0.001)\\
$Year_{2022} \times December$  &       0.047***&     (0.001)\\
Constant & 9.477*** & (0.037)\\
\hline
Cluster        &        400         &            \\
Observations   &        9600        &            \\
\hline\hline
\end{tabular}
\begin{small} \textit{Notes:} Region-level regression estimates of the natural logarithm of the number of minijobs on year times month dummies. Heteroskedasticity-robust standard errors in parentheses: * = p$<$0.10. ** = p$<$0.05. *** = p$<$0.01. \\ \textit{Source:} County-level administrative employment data of the Federal Employment Agency. 
\end{small}
\end{threeparttable}
\end{center}
\end{table}

\setcounter{figure}{0}
\setcounter{table}{0}
\renewcommand*{\thefigure}{\thesection\arabic{figure}}
\renewcommand*{\thetable}{\thesection\arabic{table}}
\clearpage
\section{Heterogeneity by job type, full graphical illustration}
\label{app:heterogeneities}

\begin{figure}[ht!]
\centering
\caption{Hourly and monthly wage effect for regular employees and minijobbers}\label{fig:12_minijob_wage_full}
\includegraphics[width=0.98\textwidth]{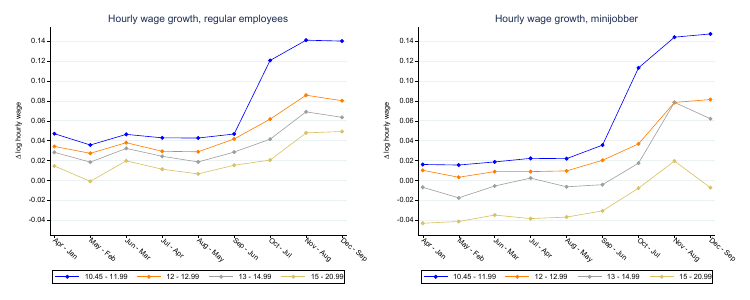}
\includegraphics[width=0.98\textwidth]{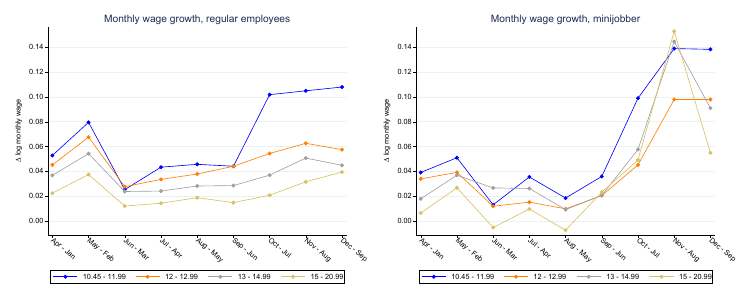}
\begin{justify}
\begin{small}{\textit{Notes:} Separate estimations and predictions for regular social security employees (on the left) and minijobbers (on the right). The upper part presents the development of hourly wage growth and the lower part presents the development of monthly wage growth. Each point is a prediction using the estimates of equation~(\ref{eq:didid}) for the respective group $g$ and month $m$, i.e., each point is an estimate of $E[\ln y_{g,m+3} - \ln y_{g,m}]$. \\
\textit{Source:} Earnings Survey, monthly panel January-December 2022, public sector excluded. }
\end{small}
\end{justify}
\end{figure}

\begin{figure}[ht!]
\centering
\caption{Working hours and employment retention effect for regular employees and minijobbers}\label{fig:12_minijob_employment_full}
\includegraphics[width=0.98\textwidth]{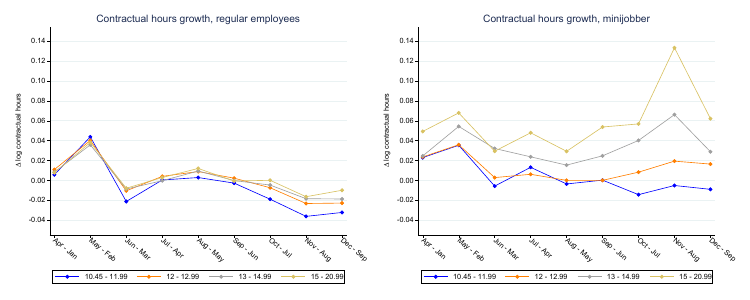}
\includegraphics[width=0.98\textwidth]{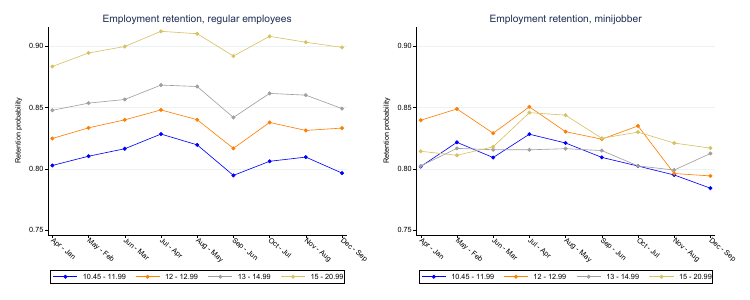}
\begin{justify}
\begin{small} {\textit{Notes:} Separate estimations and predictions for regular social security employees (on the left) and minijobbers (on the right). The upper part presents the development of monthly working hours growth and the lower part presents the development of employment retention. For monthly working hours, each point is a prediction using the estimates of equation~(\ref{eq:didid}) for the respective group $g$ and month $m$, i.e., each point is an estimate of $E[\ln y_{g,m+3} - \ln y_{g,m}]$. For employment retention, each point is a prediction using estimates of equation~(\ref{eq:did_Ret}) for the respective group $g$ and month $m$, i.e., each point is an estimate of $E[\ln E_{g,m+3}]$. \\
\textit{Source:} Earnings Survey, monthly panel January-December 2022, public sector excluded. }
\end{small}
\end{justify}
\end{figure}

\end{appendices}

\end{document}